\documentclass[12pt,preprint]{aastex62}
\usepackage{booktabs}
\usepackage{float}
\def\lea{\mathrel{<\kern-1.0em\lower0.9ex\hbox{$\sim$}}}
\def\gea{\mathrel{>\kern-1.0em\lower0.9ex\hbox{$\sim$}}}

\graphicspath{{./}{}}
\received{}
\revised{}
\accepted{}
\submitjournal{ApJ}

\shorttitle{GMC and Cluster Mass Functions in Six Nearby Galaxies}
\shortauthors{Mok et al.}

\def\aap{A\&A}

\def\apjl{ApJ}

\def\apjs{ApJS}

\def\lea{\mathrel{<\kern-1.0em\lower0.9ex\hbox{$\sim$}}}
\def\gea{\mathrel{>\kern-1.0em\lower0.9ex\hbox{$\sim$}}}

\begin{document}
\title{Mass Functions of Giant Molecular Clouds and Young Star Clusters in Six Nearby Galaxies}
\correspondingauthor{Angus Mok}
\email{mok.angus@gmail.com}
\author[0000-0001-7413-7534]{Angus Mok}
\affil{Department of Physics \& Astronomy, The University of Toledo, Toledo, OH 43606, USA}
\author{Rupali Chandar}
\affil{Department of Physics \& Astronomy, The University of Toledo, Toledo, OH 43606, USA}
\author{S. Michael Fall}
\affil{Space Telescope Science Institute, Baltimore, MD 21218, USA}
\begin{abstract}
We compare the mass functions of young star clusters (ages $\leq 10$~Myr) and giant molecular clouds (GMCs) in six galaxies that cover a large range in mass, metallicity, and star formation rate (LMC, M83, M51, NGC 3627, the Antennae, and NGC 3256). We perform maximum-likelihood fits of the Schechter function, $\psi(M) = dN/dM \propto M^{\beta} \exp(-M/M_*)$, to both populations. We find that most of the GMC and cluster mass functions in our sample are consistent with a pure power-law distribution ($M_* \rightarrow \infty$). M51 is the only galaxy that shows some evidence for an upper cutoff ($M_*$) in both populations. Therefore, physical upper mass cutoffs in populations of both GMCs and clusters may be the exception rather than the rule. When we perform power-law fits, we find a range of indices $\beta_{\rm PL}=-2.3\pm0.3$ for our GMC sample and $\beta_{\rm PL}=-2.0\pm0.3$ for the cluster sample. This result, that $\beta_{\rm Clusters} \approx \beta_{\rm GMC} \approx -2$, is consistent with theoretical predictions for cluster formation and suggests that the star-formation efficiency is largely independent of mass in the GMCs.
\end{abstract} 
\section{Introduction}\label{sec-introduction}
\par
A comparison of the mass functions of giant molecular clouds (GMCs) and young stellar clusters (with ages $\tau \lea 10$~Myr) in nearby galaxies provides important clues to the star formation process. Clusters form in the densest parts, the clumps, within GMCs (see \citet{Krumholz19} for a recent review). Even as they form, clusters begin to lose mass through feedback from massive stars. This feedback eventually removes the remaining gas, thereby limiting the star formation efficiency (SFE) and setting the shape of the cluster mass function. In a first approximation, the mass functions of both GMCs and clusters can be described by a simple power-law, $\psi(M) = dN/dM \propto M^{\beta}$. Some recent works suggest that these mass functions may have a truncation or downturn at the upper end. This can be represented by a \citet{Schechter76} function, $\psi(M) \propto M^{\beta}~\mbox{exp}(-M/M_*)$, i.e. a power-law with an exponential cutoff at $M_*$.
\par
It is important to distinguish here between {\em statistical} and {\em physical} cutoffs in the mass function. All samples have an apparent or statistical upper ``cutoff'', simply because they run out of objects (GMCs or clusters). In the case of a pure power-law with $\beta\approx-2$, this maximum cluster mass is expected to scale approximately linearly with the total number of objects due to sampling statistics, so that galaxies with more GMCs and clusters have higher apparent cutoff masses than galaxies with fewer GMCs and clusters. To date, published cutoff masses for cluster populations are largely consistent with this expected size-of-sample effect \citep[e.g.][]{Mok19}. In this work, we assess whether or not the data show a physical, i.e. an exponential-like, downturn at the upper end of the mass function {\em that is not simply the result of sampling statistics}.
\par
It is well established that the mass functions of young star clusters have $\beta \approx-2.0\pm 0.2$ \citep[e.g.][]{Zhang99, Fall12, Chandar17, Krumholz19}. Currently, the shapes of the mass functions of GMCs are less well determined, but theory and simulations suggest they should also have $\beta\approx-2$. \citep[e.g.][]{Elmegreen96, Fleck96, Wada00, Guszejnov18}. Observational work on GMC populations in nearby galaxies has found values of $\beta$ ranging from $-1.5$ in the inner Milky Way disk to $-2.9$ in M33 \citep{Rosolowsky05}, with results for other galaxies somewhere between these two extremes \citep[e.g.][]{Blitz07}. At least part of the reason for the large variation in published results is likely the use of different fitting methods, observational techniques, mass ranges studied, selection criteria, and other assumptions made in the analysis. More recent observational results have found a smaller range of $\beta$ for populations of GMCs in those same galaxies, with values such as $-1.8$ in the inner Milky Way \citep{Rice16} and $-2.0$ in M33 \citep{Gratier12}. However, it remains unclear if the power-law index $\beta$ for the observed mass functions of GMCs varies significantly among galaxies, or if there is a near-universal value (as found for cluster populations). Another open question is whether GMC mass functions are flatter, steeper, or similar to those of the young clusters that form in the same galaxy. 
\par
It also remains uncertain whether the upper cutoffs in the mass functions of young clusters and GMCs are predominantly statistical or physical. For {\em cluster} populations, cutoffs have been claimed in NGC~4041 \citep{Konstantopoulos03}, M83 \citep{Bastian12, Adamo15}, M31 \citep{Johnson17}, M51 \citep{Messa18a}, and the Antennae \citep{Jordan07}. On the other hand, \citet{Mok19} applied a uniform maximum-likelihood fitting procedure to a sample of young clusters in eight nearby galaxies, including M83, M51, and the Antennae, and found that the majority of the galaxies do not show evidence for a physical cutoff, but are consistent with the expectations of a statistical cutoff. \citet{Whitmore20} reached a similar conclusion about physical cutoffs for clusters in NGC~4449, as did \citet{Cook19} for a composite of 17 dwarf galaxies studied as part of the LEGUS project. For {\em GMC} populations, evidence of a cutoff $M_*$ at the $>3\sigma$ level has been claimed for some galaxies (e.g., M33 \citep{Rosolowsky07}, NGC~4256 \citep{Utomo15}), with weak ($2-3\sigma$) evidence (e.g., M51 \citep{Colombo14}, NGC~300 \citep{Faesi18}, NGC 6946 \citep{Wu17}), or no evidence found in others (e.g. in outer disk of the Milky Way \citep{Rice16} and the LMC \citep{Wong11}). One of the main goals of this work is to use the same fitting method for GMC and cluster samples. We will use a uniform procedure for both populations to establish their relation to one another and determine any galaxy-to-galaxy variations.
\par
The relation between the shapes of the GMC and cluster mass functions is dictated by early stellar feedback mechanisms. \citet[][]{Fall10}, hereafter FKM, analytically derived simple relations between the power-law indicies of gas-dominated protoclusters and the resulting stellar clusters in the limiting cases of energy- and momentum-driven feedback, which correspond to the minimum and maximum radiative losses inside protoclusters (clumps and GMCs). These relations in turn depend on the relation between the radii and masses of the protoclusters, $R_h \propto M^{\alpha}$. Observations in the Milky Way and LMC indicate $\alpha\approx0.5$, i.e. roughly constant mean surface density, for both GMCs \citep{Larson81, Blitz07} and clumps \citep[][hereafter FKM]{Wu10, Wong19}. For $\alpha\approx0.5$ and typical values of $\beta_{\rm GMC} \approx -2$, the FKM model predicts $\beta_{\rm GMC} \approx \beta_{\rm Clusters} \approx-2$. This is because the SFE is independent of the initial masses of the protoclusters and dependent mainly on their surface density ($\Sigma$). Recent hydrodynamical simulations of star-formation inside molecular clouds of various masses agree with the predicted SFE from this analytical model \citep[e.g.][]{Kim18, Grudic18}.
\par
It has also been suggested that the SFE can be measured using the ratio between the upper-mass cutoff or truncation mass $M_*$ in the GMC and cluster populations \citep[e.g.][]{Gieles06, Kruijssen14}. This method has been recently applied in M51, with $M_{\rm *,Clusters}/M_{\rm *,GMC}$ estimated to be $\sim 1\%$ \citep{Messa18b}. However, this method is only appropriate in cases where a physical (not just a statistical) cutoff is detected with high confidence in both populations.
\par
In this work, we uniformly apply the robust maximum-likelihood fitting method developed in \citet{Mok19} to the GMC and cluster populations in 6 nearby galaxies (LMC, M83, M51, NGC 3627, Antennae, and NGC 3256), selected to have high-quality catalogs for both populations. The galaxies span a wide range in mass, morphology, and star formation rate (SFR). The rest of this paper is organized as follows. In Section~\ref{sec-gmc}, we present the GMC catalogs and their mass functions. In Section~\ref{sec-clusters}, we present the cluster catalogs and their mass functions, including a new catalog for NGC~3627. In Section~\ref{sec-maxlikelihood}, we present the results of maximum-likelihood fits of a Schechter function and power law to the cluster and GMC masses. In Sections~\ref{sec-discussion} and \ref{sec-conclusions}, we discuss and summarize the main implications of our results.
\begin{figure*}[ht]
	\centering
	\includegraphics[width=7.5cm]{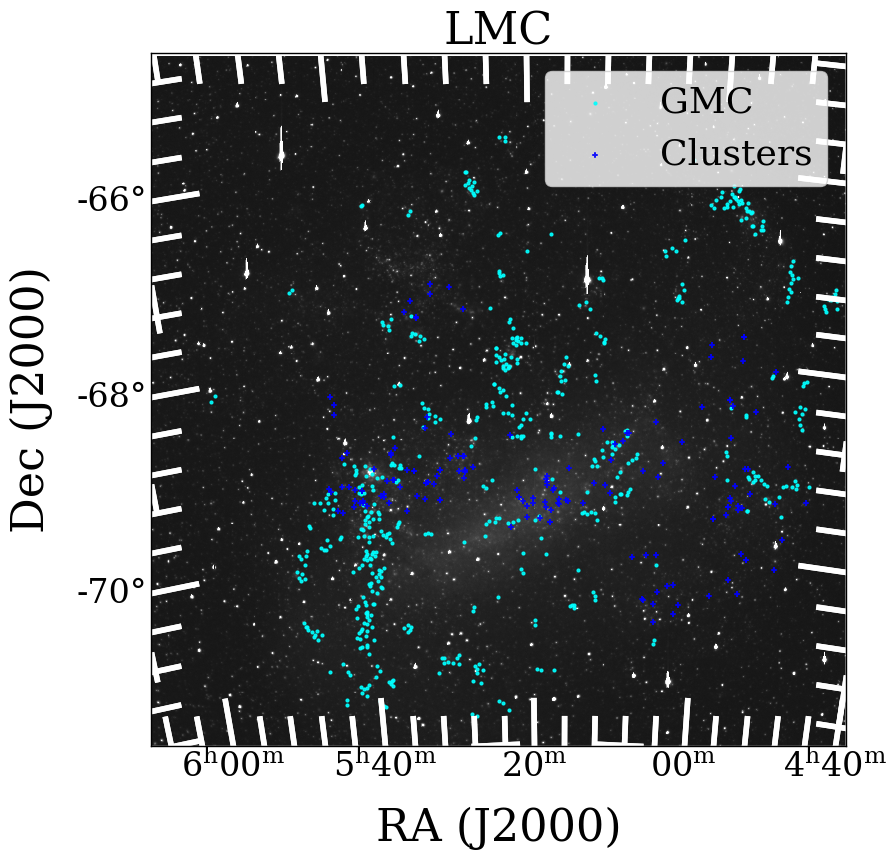}
	\includegraphics[width=7.5cm]{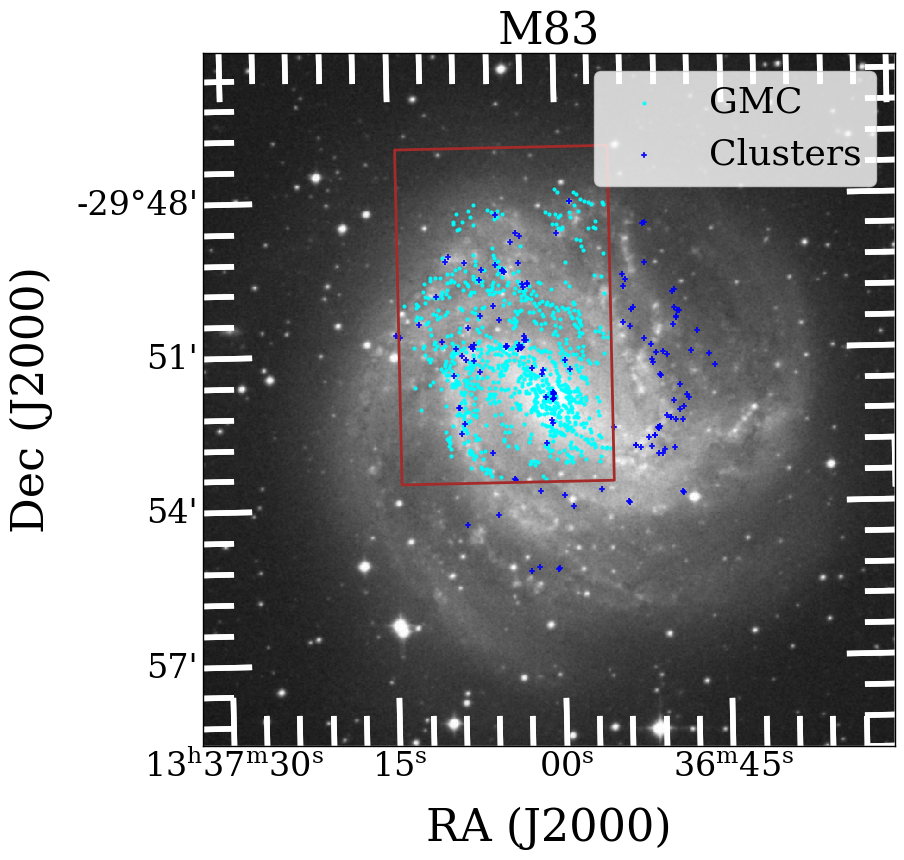}
	\includegraphics[width=7.5cm]{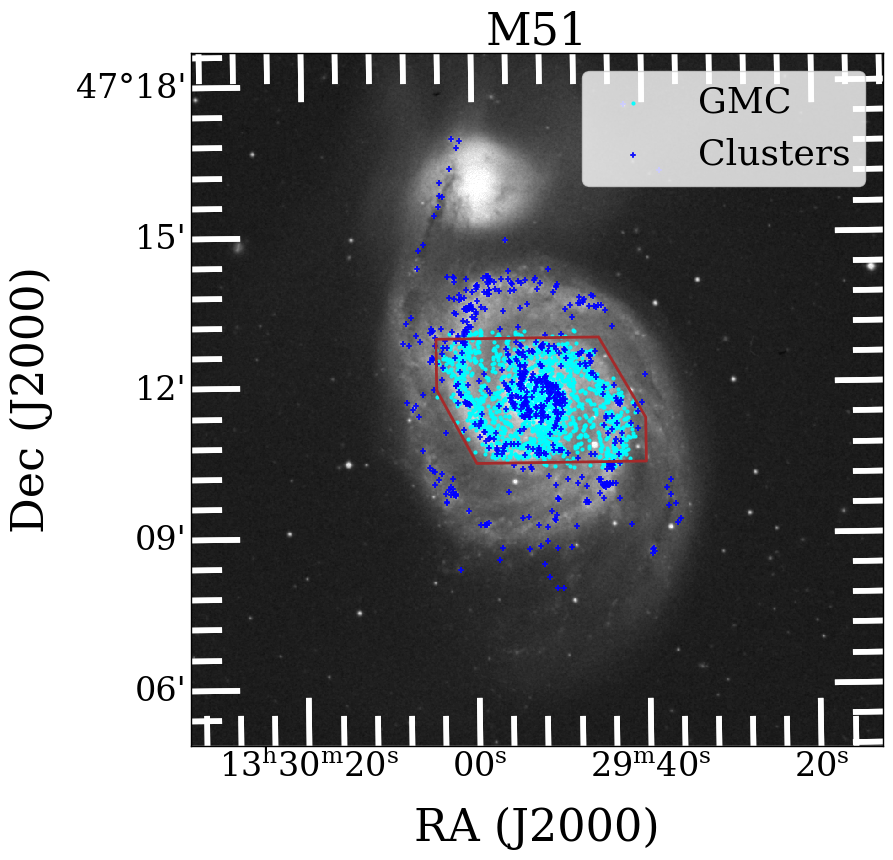}
	\includegraphics[width=7.5cm]{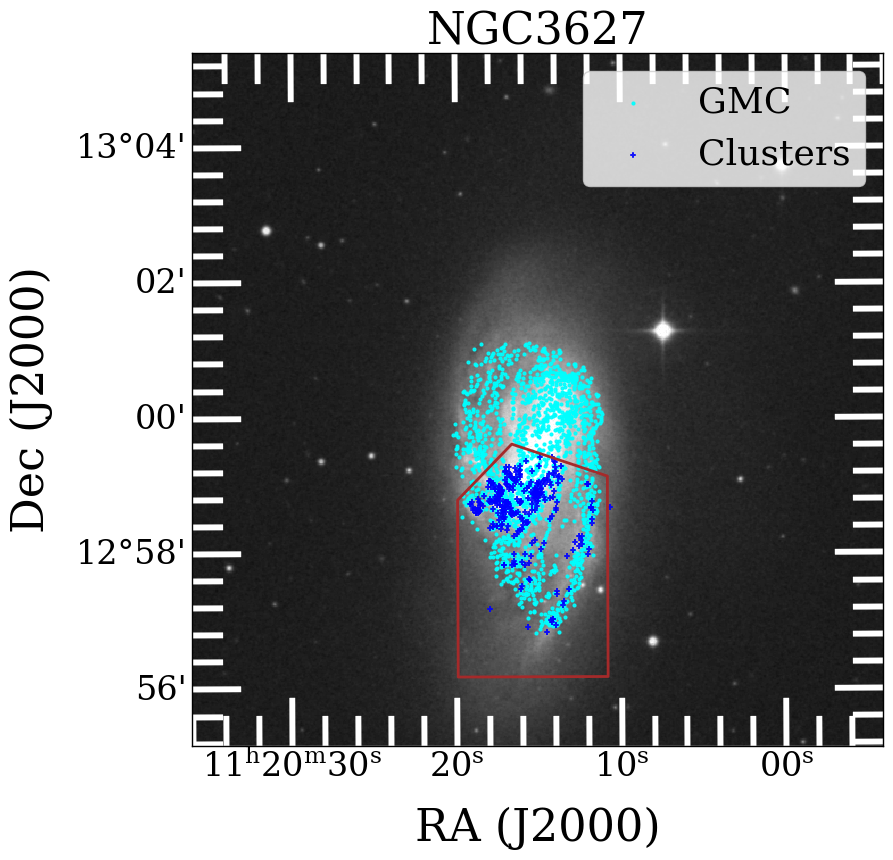}
	\includegraphics[width=7.5cm]{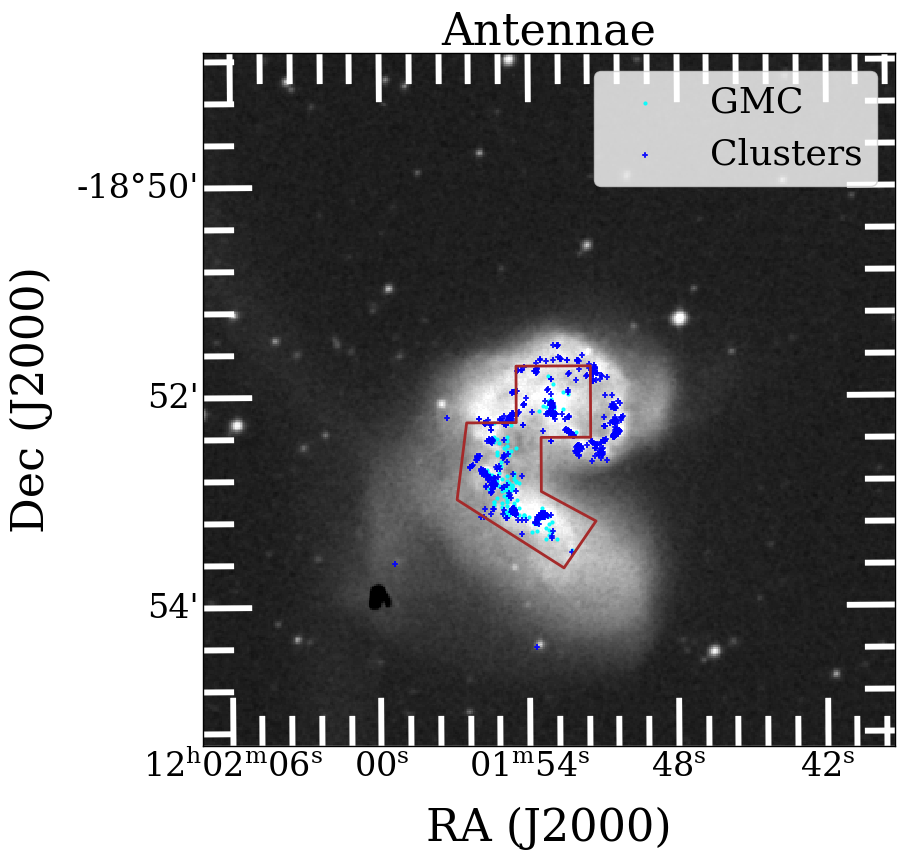}
	\includegraphics[width=7.5cm]{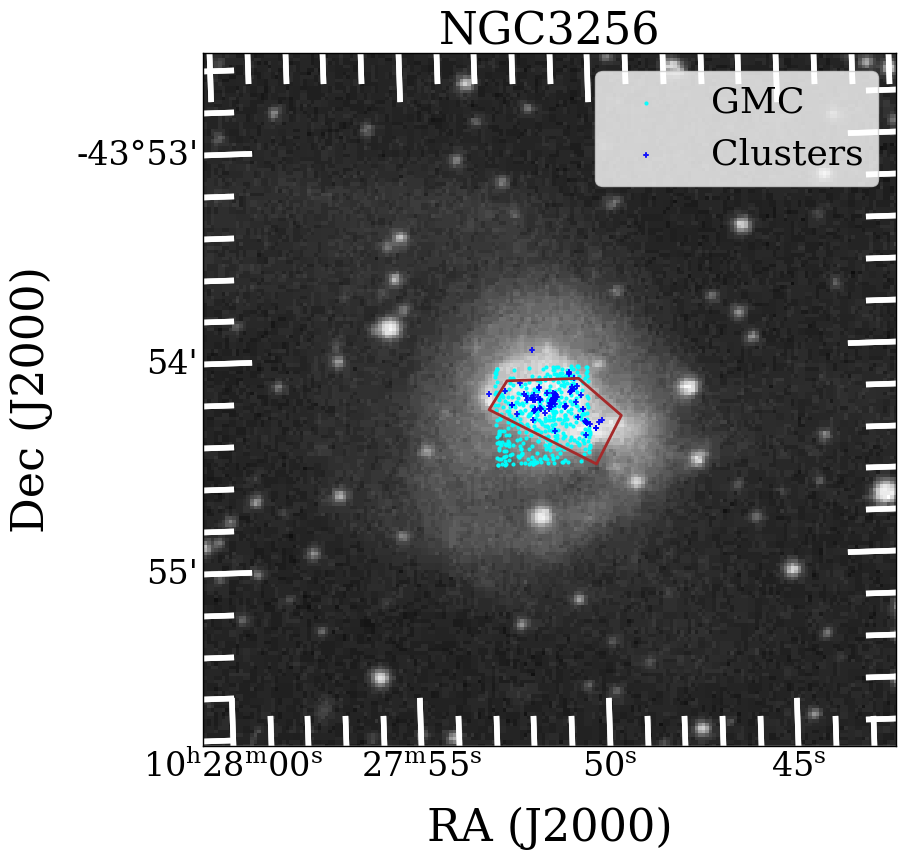}
	\caption{Ground-based images of our galaxy sample, LMC, M83, M51, NGC 3627, the Antennae, and NGC 3256 showing the locations of young star clusters (blue) and GMCs (green). The brown outline shows the area in common between the cluster and GMC catalogs. The images for five galaxies are from the Digital Sky Survey, while the LMC image is retrieved from Karl D. Gordon's website: \url{http://dirty.as.arizona.edu/~kgordon/research/mc/lmc_optical.html}.}\label{fig-galreg}
\end{figure*} 
\begin{table*}[ht]
	\caption{Summary of the Basic Properties of our Galaxy Sample}\label{tab-galprop}
	\centering
	\begin{tabular}{lcccccccc}
		Galaxy & Distance & SFR & N$_{\rm Clusters}$ \\
		- & [Mpc] & [$M_{\odot}$ yr$^{-1}$] & - \\ 
		\hline
		LMC & 0.050 & 0.25 & 931 \\ 
		M83 & 4.5 & 2.65 & 3177 \\
		M51 & 7.6 & 3.20 & 3812 \\
		NGC 3627 & 10.1 & 4.89 & 742 \\
		Antennae & 22 & 20 & $>10,000$ \\ 
		NGC 3256 & 36 & 50 & 505 \\
		\hline
	\end{tabular}
\end{table*}
\section{GMCV=s}\label{sec-gmc}
\par
In this section, we present catalogs of the GMCs in our six target galaxies. Ground-based optical images of each galaxy are shown in Figure~\ref{fig-galreg}. Some basic properties of the galaxies, such as their distance and SFR, are summarized in Table~\ref{tab-galprop}. Four out of six of our galaxies have published GMC catalogs (LMC, M83, M51, and the Antennae), which we summarize in Section~\ref{subsec-gmc-catalog}. We present new GMC catalogs for NGC~3627 and NGC~3256 based on archival Atacama Large Millimeter/submillimeter Array (ALMA) observations in Section~\ref{subsec-gmc-newcatalog}. The coverage for the molecular gas surveys is also shown in Figure~\ref{fig-galreg}, and overlaps significantly with the available cluster catalogs. Important details about the observations, such as the observed CO transition and the beam size (in arcsec and parsecs) are listed in Table~\ref{tab-gmcobs}.
\subsection{Previous Catalogs}\label{subsec-gmc-catalog}
\par
Below, we summarize some basic information about the published GMC catalogs in the LMC, M83, M51, and the Antennae:
\begin{itemize}
\item
{\bf LMC:} A catalog of 543 GMCs was published by \citet{Wong11}, based on CO(1-0) observations taken with the ATNF Mopra Telescope. They assume a value for the CO-to-H$_2$ ($\alpha_{\rm CO}$) conversion factor of 4.8 M$_\odot$ (K km s$^{-1}$ pc$^2$)$^{-1}$. The coverage is quite piecemeal, and the proximity of the LMC gives this catalog significantly higher physical resolution compared with the others in our sample.
\item
{\bf M83:} A catalog of 873 GMCs was published by \citet{Freeman17}, based on ALMA CO(1-0) observations from project 2012.1.00762.S (PI: A. Hirota). They assume a $\alpha_{\rm CO}$ value of 4.35 M$_\odot$ (K km s$^{-1}$ pc$^2$)$^{-1}$. The catalog covers the nuclear starburst and parts of the northern region.f
\item 
{\bf M51:} A catalog of 1507 GMCs was published by \citet{Colombo14}, based on IRAM CO(1-0) observations taken as part of the PAWS survey \citep{Schinnerer13}. They assume a value for $\alpha_{\rm CO}$ of 4.4 M$_\odot$ (K km s$^{-1}$ pc$^2$)$^{-1}$. The survey covers the center and both inner spiral arms of the galaxy, but not the outer arms.
\item
{\bf Antennae:} A catalog of 142 GMCs was published by \citet{Zaragoza14}, based on ALMA CO(3-2) observations from the ALMA science verification program and project 2011.0.00876.S (PI: B. Whitmore). They assume a value for the CO-to-H$_2$ ($\alpha_{\rm CO}$) conversion factor of 4.8 M$_\odot$ (K km s$^{-1}$ pc$^2$)$^{-1}$, and a line ratio between CO(3-2) and CO(1-0) of 1.8. Coverage is restricted mostly to the overlap region and both nuclei of this merging system.
\end{itemize}
\par 
From the published \citet{Colombo14} catalog, the mean uncertainty in the flux measurements of GMCs in M51 (above the adopted completeness limit) is close to $\sim0.3$ dex. We will adopt this as a fiducial value for the errors in the GMC masses.
\subsection{New Catalogs}\label{subsec-gmc-newcatalog}
\par
\begin{figure}[ht]
	\centering
	\includegraphics[width=8.5cm]{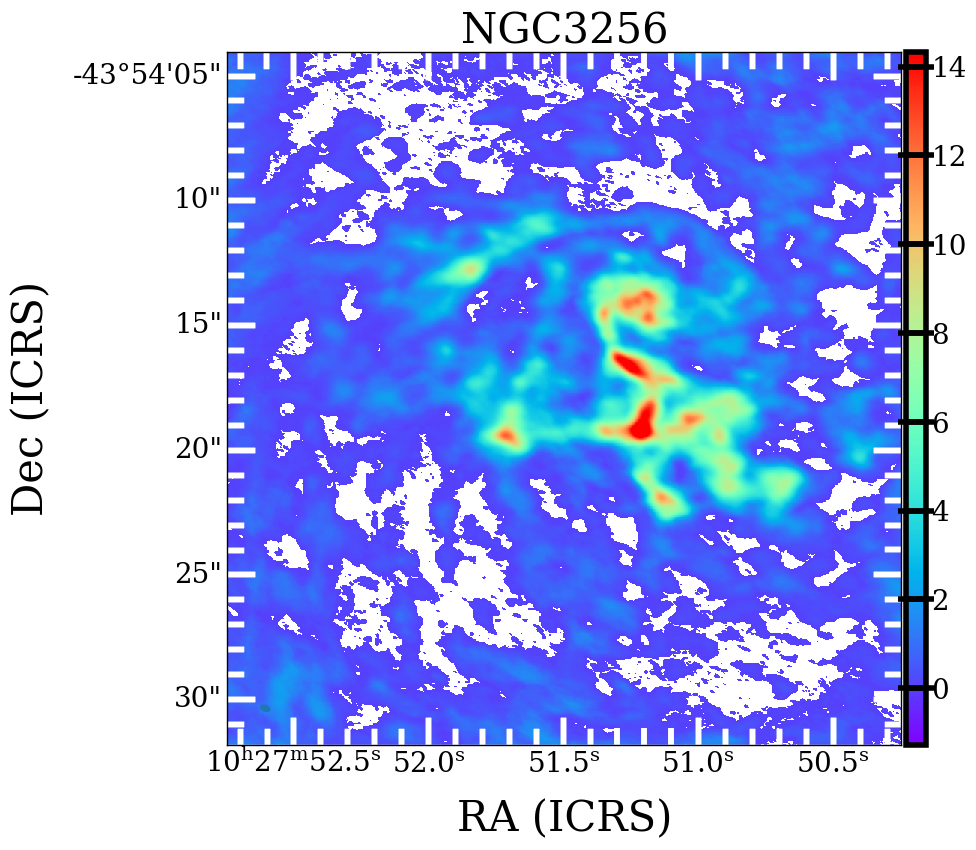}
	\includegraphics[width=8.5cm]{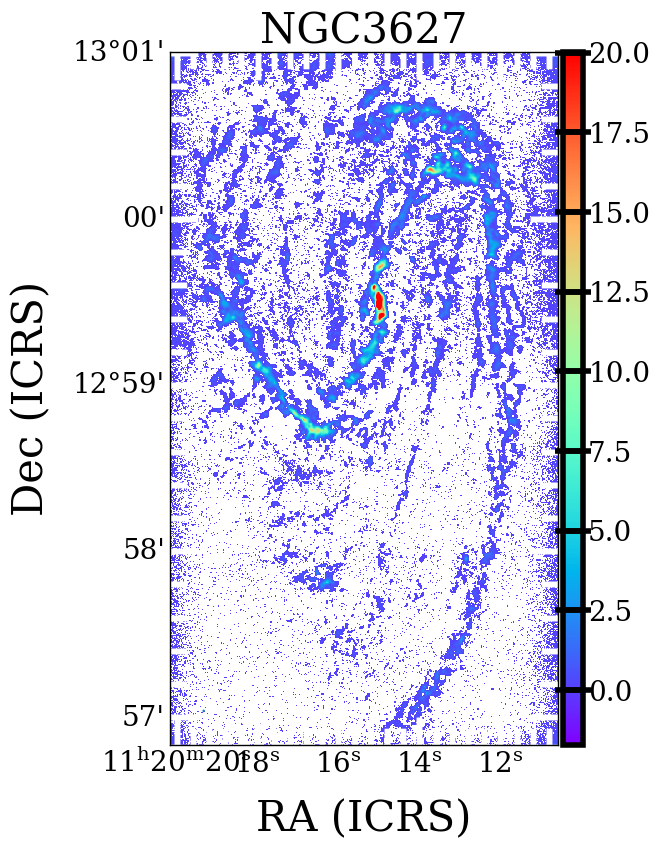}
	\caption{Integrated intensity (moment-zero) maps for the molecular gas ALMA datasets used for NGC 3256 (left) and NGC 3627 (right). The plots show the integrated CO(2-1) line intensity in units of Jy km s$^{-1}$ bm$^{-1}$.}\label{fig-gmcmommap}
\end{figure}

We produce new catalogs of GMCs for NGC~3627 and NGC~3256, since no published catalogs currently exist for these galaxies. After downloading the pipeline calibrated datasets for NGC~3627 (project 2015.1.00956.S PI: A. Leroy) and NGC~3256 (project 2015.1.00714.S, PI: K. Silwa) from the ALMA archive, we perform our data reduction and imaging using the CASA software package \citep{CASA}. We first perform continuum subtraction for the individual interferometric datasets and then combine the results from the 12-meter and 7-meter (ACA) arrays for each galaxy. To image the dataset, we use the non-interactive cleaning process in the {\sc tclean} routine. We select the area of interest in the two galaxies, using the line-free regions in the datacube to estimate the noise, and then run the {\sc tclean} process until it reaches a clean threshold of twice the measured rms noise. To obtain the final images for NGC 3627, we also run the feathering routine from CASA to incorporate the zero-spacing information from the available single-dish total power observations. The resulting moment zero (total intensity) maps are presented in Figure~\ref{fig-gmcmommap}.
\par 
We use the publicly available {\sc CPROPSTOO} code\footnote{http://github.com/akleroy/cpropstoo} to detect GMCs \citep{Rosolowsky06}. The routine selects objects that are detected at the $\geq 5\sigma$ level in two adjacent channels, where $\sigma$ is determined from signal-free regions. These regions are then grown to include all pixels with greater than 2$\sigma$ emission. Our criteria are more strict than the corresponding 2.5$\sigma$ and 1$\sigma$ thresholds, respectively, used by \citet{Freeman17}, which leads to fewer detections at the faint end. The difference in methodology has little impact on the resulting mass functions, since those low mass GMCs found in the \citet{Freeman17} catalog are not used in the fitting process (discussed further in Section~\ref{subsec-gmc-massfunctions}). We also adopt standard values for other {\sc CPROPSTOO} parameters, such as $\Delta=2$ in SNR units (which sets the required contrast) and a minimum cloud size of 25 pixels. Tests show that varying the adopted value of $\Delta$ (between 1.0 to 3.0) has only a small effect on our conclusions.
\par
For each object, we convert the resulting CO luminosity to H$_2$ mass by assuming a standard CO-to-H$_2$ conversion factor of 4.35 M$_\odot$ (K km s$^{-1}$ pc$^2$)$^{-1}$ and a line ratio between CO(2-1) and CO(1-0) of 0.8 \citep[e.g.][]{Leroy09}. We also scale the other catalogs to this CO-to-H$_2$ conversion factor. While this value is appropriate for normal, star-forming galaxies, significantly lower values (such as 0.8 M$_\odot$ (K km s$^{-1}$ pc$^2$)$^{-1}$) have been suggested for starbursts and Ultra-Luminous Infra-Red Galaxies (ULIRGs) \citep{Bolatto13}. The two interacting galaxies in our sample, the Antennae and NGC 3256, are both LIRGs and have high rates of star formation. There are also uncertainties in the assumed value of the line ratio used to convert from CO(2-1) to CO(1-0). Previous works have found a range from 0.6 to 1.0 in nearby galaxies \citep[e.g.][]{Leroy09}. For the Antennae observations, \citet{Zaragoza14} assume a value of 0.8 to convert from CO(3-2) to CO(1-0), motivated by previous observations \citep{Ueda12}, and thus we continue to use this value here. The uncertainties associated with the assumed CO-to-H$_2$ conversion factor, line ratio, and distance to the galaxy itself do not affect the {\em shape} of the mass function or the main conclusions of this paper, but will affect the value of $M_*$.
\subsection{Mass Functions}\label{subsec-gmc-massfunctions}
\par
\begin{figure*}[ht]
	\centering
	\includegraphics[width=16.5cm]{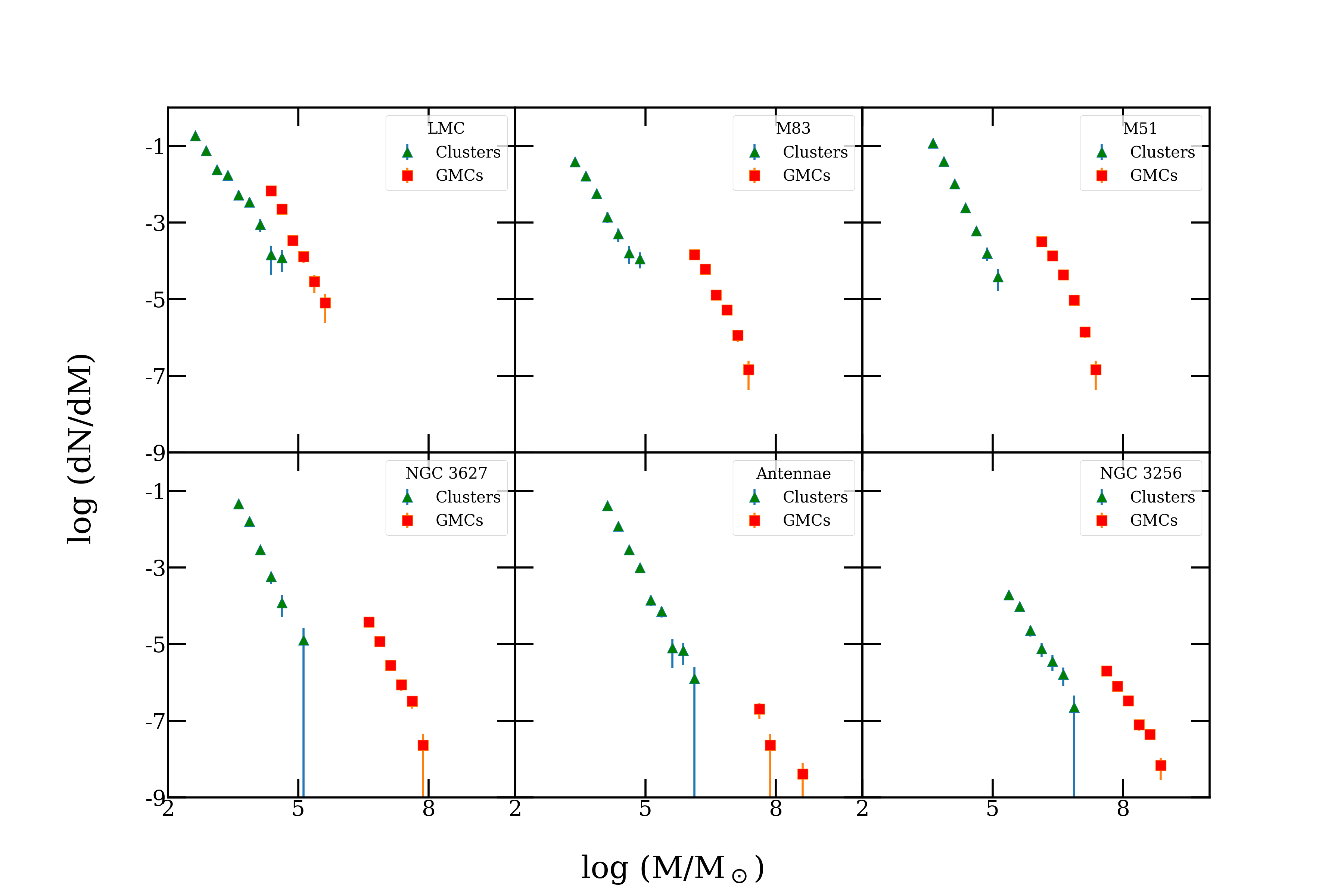}
	\caption{The mass functions of GMCs (red squares) and young ($<10$~Myr) clusters (blue triangles) plotted using equal logarithmic bins. We only show data above the completeness limit. Note that the binned mass functions shown here are only for visual purposes and are not used in the maximum-likelihood fitting.}\label{fig-scgmccompbin}
\end{figure*}

We present binned versions of the GMC mass functions in Figure~\ref{fig-scgmccompbin} (red squares). Each distribution appears to follow a power-law reasonably well when plotted using equal logarithmic bins, although this binning method can hide weak features at the ends of the distribution. We also find that just as for clusters, the maximum GMC masses approximately scale with their total numbers, as expected from the size of sample effect. In Section~\ref{sec-maxlikelihood}, we will use our maximum-likelihood method to fit a Schechter function to the GMC masses.
\par
\begin{figure}[ht]
	\centering
	\includegraphics[width = 16.5 cm]{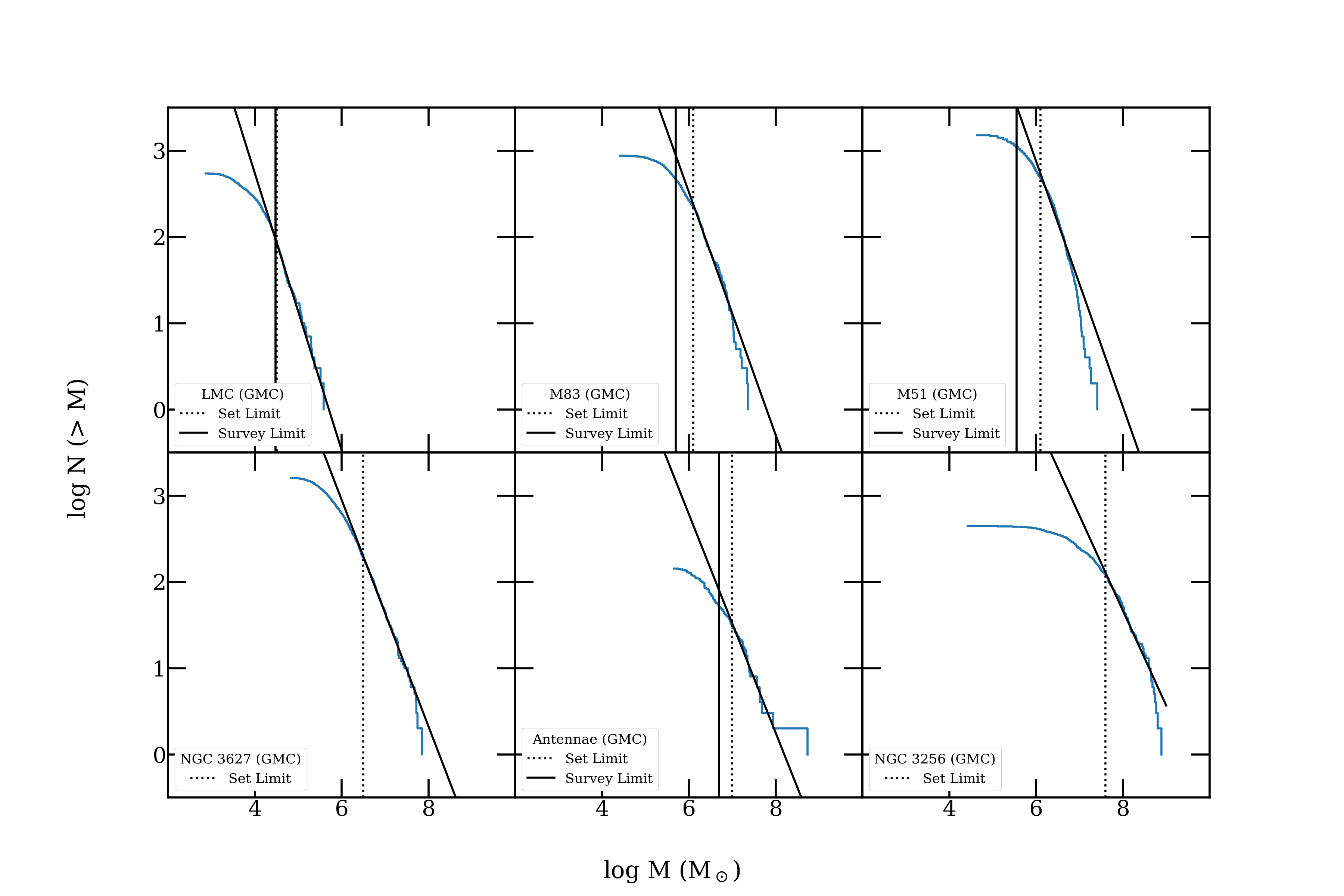}
	\caption{Cumulative distributions of the GMC masses are plotted as the solid blue lines. A power-law fit (black solid line) is added to help guide the eye. The completeness limits determined as described in Section~\ref{subsec-gmc-massfunctions} are indicated by the dotted vertical lines, and the published completeness limits as solid lines, when available.}\label{fig-completeness}
\end{figure}

It is important to establish a lower mass limit above which each catalog is complete. The fitting procedure can erroneously find a cutoff at higher masses even when there isn't one, if sources below the completeness limit (where the distribution flattens) are included. Differences in physical resolution can also potentially affect mass estimates at the lower end of each survey. To minimize the impact of incompleteness and differences in resolution, we apply a uniform procedure (demonstrated in Figure~\ref{fig-completeness}) to each GMC and cluster catalog to establish a lower mass limit. This figure shows that the cumulative mass distribution for each GMC catalog follows a power law at the upper end, but eventually flattens toward lower masses. We assume that this flattening is due to incompleteness, rather than to a physical effect, just as we have done previously for cluster catalogs \citep[e.g.][]{Chandar17, Mok19}. We set the completeness limit $M_{lim}$ for each catalog at the mass where the distribution begins to flatten noticeably, represented by the dotted lines in Figure~\ref{fig-completeness}, and listed in the second column of Table~\ref{tab-likefitgmc}. We show the published completeness limits for comparison, when available, as a solid vertical line. Our method provides stricter lower mass limits than those published for M83, M51, and the Antennae catalogs.
\par
While the galaxies in our sample have a large range of distances, the observations used here (with the exception of the LMC) all have a physical resolution that is within $\sim50$\% of 50~pc. We tested the potential impact that this range of resolution might have on our results by convolving the NGC~3627 data to match the physical scale of the NGC~3256 catalog, then rerunning it through our detection and fitting software, and found only a small effect on the maximum-likelihood results (described in Section~\ref{sec-maxlikelihood}). Finally, we note that the GMC dataset for the LMC is quite different from the others, since it has significantly higher physical resolution and piecemeal coverage over the galaxy. We present results in this paper for the LMC, but we note that they may not be directly comparable to those from the other five galaxies in the sample.
\begin{table*}[ht]
	\caption{Summary of the Molecular Gas Observations and GMC Catalogues}\label{tab-gmcobs}
	\centering
	{
	\begin{tabular}{lccccc}
		Galaxy & Line & $\alpha_{\rm CO}$ & Line Ratio & Beam (angular) & Beam (physical) \\
		 - & - & [M$_\odot$ (K km s$^{-1}$ pc$^2$)$^{-1}$] & & [arcsec] & [pc] \\ 
	 \hline
		\multicolumn{6}{c}{Previous Work} \\
		LMC$^1$ & CO(1-0) & 4.35 & - & 45 & 11 \\
		M51$^2$ & CO(1-0) & 4.35 & - & 1.16$\times$0.97 & 53$\times$36 \\
		M83$^3$ & CO(1-0) & 4.35 & - & 1.43$\times$0.83 & 29$\times$18 \\
		Antennae$^4$ & CO(3-2) & 4.35 & 1.8 & 0.6$\times$1.1 & 64$\times$117\\
		 & & & & 0.4$\times$0.7 (C0) & 41$\times$62 (C0)\\
		\hline
		\multicolumn{6}{c}{This Work} \\
		NGC 3627$^5$ & CO(2-1) & 4.35 & 0.8 & 0.95$\times$0.88 & 46$\times$43 \\
		NGC 3256$^6$ & CO(2-1) & 4.35 & 0.8 & 0.45$\times$0.28 & 78$\times$49 \\	
		\hline
	\end{tabular}
	}
	{\raggedleft
	\begin{tabular}{l}
	 \noindent
		$^1$ \citet{Wong11}, $^2$ \citet{Colombo14}, $^3$ \citet{Freeman17}, $^4$ \citet{Zaragoza14}, \\ $^5$ 2015.1.00956.S (PI: A. Leroy), $^6$ 2015.1.00714.S (PI: K. Silwa) \\
		Note: C0 indicate properties from the Cycle-0 observations for the Antennae
	 \end{tabular}
	 }
\end{table*}
\section{Star Clusters}\label{sec-clusters}
\subsection{Previous Catalogs and Mass Functions}\label{subsec-clusters-catalog}
\par
All of the cluster catalogs used in this work, with the exception of NGC~3627, have been collected from the literature, and the resulting mass functions published in \citet{Chandar15, Chandar17}. Here we summarize basic information on each catalog, including where the observations were taken, the cluster selection criteria, and the method for estimating ages and masses. 
\par
The LMC cluster catalog is presented by \citet{Hunter03}, and based on ground-based {\it UBVR} images at with the Cerro Tololo Inter-American Observatory. The M51 catalog is presented by \citet{Chandar16} based on Hubble Space Telescope (HST) {\it UBVI}H$\alpha$ observations. The M83 catalogue comes from new HST {\it UBVI}H$\alpha$ observations (Whitmore et al., in prep). The Antennae catalog is presented by \citet{Whitmore10}, and based on {\em HST} {\it UBVI}H$\alpha$ observations. The cluster catalog for NGC 3256 comes from \citet{Mulia16}, based on {\it UBVI}H$\alpha$ HST observations. Note that we correct the Antennae and M51 masses by a factor of 0.6 from the original adopted \citet{Salpeter55} initial mass function (IMF) to a \citet{Chabrier03} IMF here, which is assumed for the other catalogs.
\par
The clusters in the catalogs used here are all selected to have a higher stellar density than the local background, with no attempt made to assess whether or not they are gravitationally bound based on their morphology, as has been advocated in some works \citep[e.g.][]{Adamo17}. We believe that this is the best practice when comparing samples of clusters in galaxies with a large range of distances, since morphology cannot be evaluated in a uniform way between the closest and most distant galaxies, which could potentially lead to a distance-related bias. Furthermore, simulations have shown that it is not possible to discern if a cluster is gravitationally bound (has negative or positive energy) based solely on its morphology \citep[e.g.][]{Baumgardt07}. In \citet{Mok19}, we found that cluster catalogs in M83 and M51 created by different groups, despite having somewhat different methods and criteria, gave similar results for the shape of the mass function.
\par
The age and mass of each cluster is estimated by comparing its integrated multiband photometry with predictions from the \citet{Bruzual03} stellar population models. All galaxies studied here, except the LMC, have narrow-band measurements (H$\alpha$), which includes both stellar continuum and nebular line emission, in their fits. We found that H$\alpha$ is important in disentangling the effects of age and reddening in the broad-band measurements of clusters \citep[e.g.][]{Fall05, Whitmore20}, but also verified that the LMC cluster age estimates are robust \citep{Chandar10}. To estimate the mass of each cluster, we use the age-dependent mass-to-light ratio, the extinction-corrected $V$-band magnitude, and assume an underlying \citet{Chabrier03} IMF and the distance to each galaxy listed in Table~\ref{tab-galprop}. The fitting method typically introduces an uncertainty of 0.3 in log~M ($\approx$ factor of 2) in the masses \citep[e.g.][]{Elson88, deGrijs06, Chandar10}.
\par
We compare the mass functions of clusters younger than 10 Myr (blue triangles) in Figure~\ref{fig-scgmccompbin} with those of the GMCs in the same galaxies. While it is possible that the cluster mass functions for the most distant galaxies in our sample may suffer from biases related to crowding, previous works have found the shapes of these distributions to be fairly stable out to the distances of $\approx40$~Mpc. \citet{Randriamanakoto13} and \citet{Mulia16} tested the impact of `distance bias' by degrading images of galaxies to simulate distances similar to the two most distant galaxies used here, and found the resulting power-law index of the mass function does not differ by more than $\approx0.2$.
\subsection{New Catalog and Mass Function for NGC~3627}\label{subsec-clusters-ngc3627}
\par
Here, we present a new cluster catalog for NGC~3627 based on broad-band images taken in five filters as part of the LEGUS project (UV, UBVI; \citep{Calzetti15}), plus narrow-band H$\alpha$ photometry taken as part of H$\alpha$-LEGUS. Nearly 1600 candidate compact clusters were selected following the procedure described in \cite{Adamo17}.
\par
Photometry is performed in a 4 pixel aperture radius for all filters. No continuum subtraction is performed on the narrow-band H$\alpha$ image, so the measurements contain a combination of nebular line and stellar continuum emission. We use two different methods to determine the aperture correction. First, we determine an average aperture correction of 0.834~mag from a number of relatively isolated clusters. Second, we fitted a function to the measured aperture correction and concentration index ($C$, the difference between aperture magnitudes in 0.5 and 3~pixels; see for example \citet{Cook19}) for synthetic clusters. We find that the mass functions are quite similar for both methods of determining aperture corrections. The instrumental magnitudes are converted to the VEGAMAG photometric system by applying zero points available from the instrument page on the STScI website. We use a similar method to the one described in Section~\ref{subsec-clusters-catalog} to estimate the masses and ages of the clusters.
\par
\begin{figure*}[ht]
	\centering
	\includegraphics[width=8.5cm]{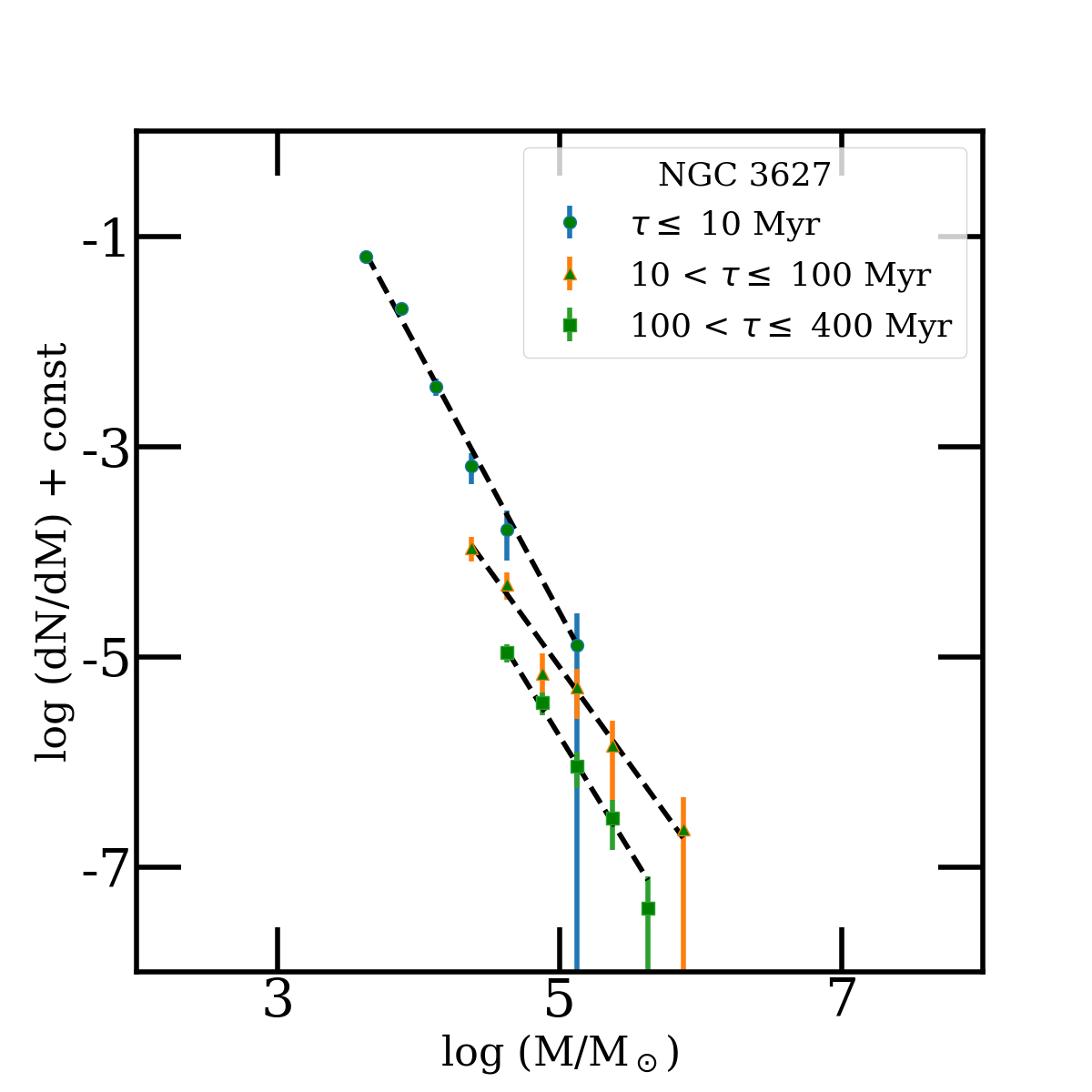}
	\caption{Mass functions of star clusters in NGC~3627 using equal logarithmic bins are plotted in three intervals of age: $\tau < 10$ Myr (circles), $\tau = 10 - 100$ Myr (triangles), and $\tau = 100 - 400$ Myr (squares). The dashed lines show the best-fit power law to each distribution, where $dN/dM \propto M^{-\beta_{\rm PL}}$. The best-fit values of $\beta_{\rm PL}$ to the binned points are: $-2.49\pm0.11$ for $\tau < 10$~Myr, $-1.86\pm0.20$ for $\tau=10-100$~Myr, and $-2.19\pm0.23$ for $\tau=100-400$~Myr clusters.}\label{fig-ngc3627}
\end{figure*}

In Figure~\ref{fig-ngc3627}, we plot the binned mass functions of clusters in NGC~3627 in three different intervals of age: $\tau < 10$ Myr, $\tau = 10 - 100$ Myr, and $\tau = 100 - 400$ Myr. The corresponding completeness limits in these age bins are $10^{3.5}$ $M_\odot$, $10^{4.2}$ $M_\odot$, and $10^{4.5}$ $M_\odot$, where the number of clusters above the limits are 289, 44, and 118 respectively. We fitted power laws to the binned distribution, with best fit indicies of $-2.49 \pm 0.11$ ($\tau < 10$ Myr), $ -1.86 \pm 0.20$ ($\tau = 10 - 100$ Myr), and $-2.19 \pm 0.23$ ($\tau = 100 - 400$ Myr) clusters. Note that in the rest of this paper, we only consider clusters with ages younger than 10 Myr.
\section{Maximum-Likelihood Fits}\label{sec-maxlikelihood}
\par
In this Section, we use the maximum-likelihood method described in \citet{Mok19} to determine the best-fit values and confidence intervals for the parameters $\beta$ and $M_*$ when fitting a Schechter function to the cluster and GMC masses above the completeness limit of each catalog. This method has the advantage of not using binned data (which can hide weak features at the ends of the distribution) or cumulative distributions (where the data points are not independent of one another). We compute the likelihood $L(\beta,M_*)=\Pi_i P_i$ as a function of $\beta$ and $M_*$, where the probability $P_i$ for each cluster is given by:
\begin{equation}
P_i = \frac{\psi(M_i)}{\int_{M_{\rm min}}^{\infty} \psi(M)dM} 
\end{equation}
and the product is over all GMCs or clusters above $M_{min}$ (see e.g. Chapter~15.2 of \citet{Mo10}). For each catalog, we set the upper integration limit in equation (1) to be 100 times the mass of the most massive cluster in that sample; our tests showed that this was sufficient for convergence in all cases. Next, we find the maximum-likelihood $L_{\rm max}$ using the \citet{Nelder65} method, and use the standard formula:
\begin{equation}
\ln L(\beta, M_*) = \ln L_{\rm max} - \frac{1}{2} \chi_p^2(k)
\end{equation}
where $\chi_p^2(k)$ is the chi-squared distribution with $k$ degrees of freedom at $p$ confidence level to determine the 1-, 2-, and 3-$\sigma$ confidence contours. We present our results for the zero error case in Section~\ref{subsec-maxlikelihood-result}, and for the case including a typical uncertainty of $\sigma (\log M) \approx0.3$ in Section~\ref{subsec-maxlikelihood-result-error}.
\subsection{Results Without Measurement Uncertainties}\label{subsec-maxlikelihood-result}
\par
\begin{figure}[ht]
	\centering
	GMCs: \\
	\includegraphics[width = 12.5 cm]{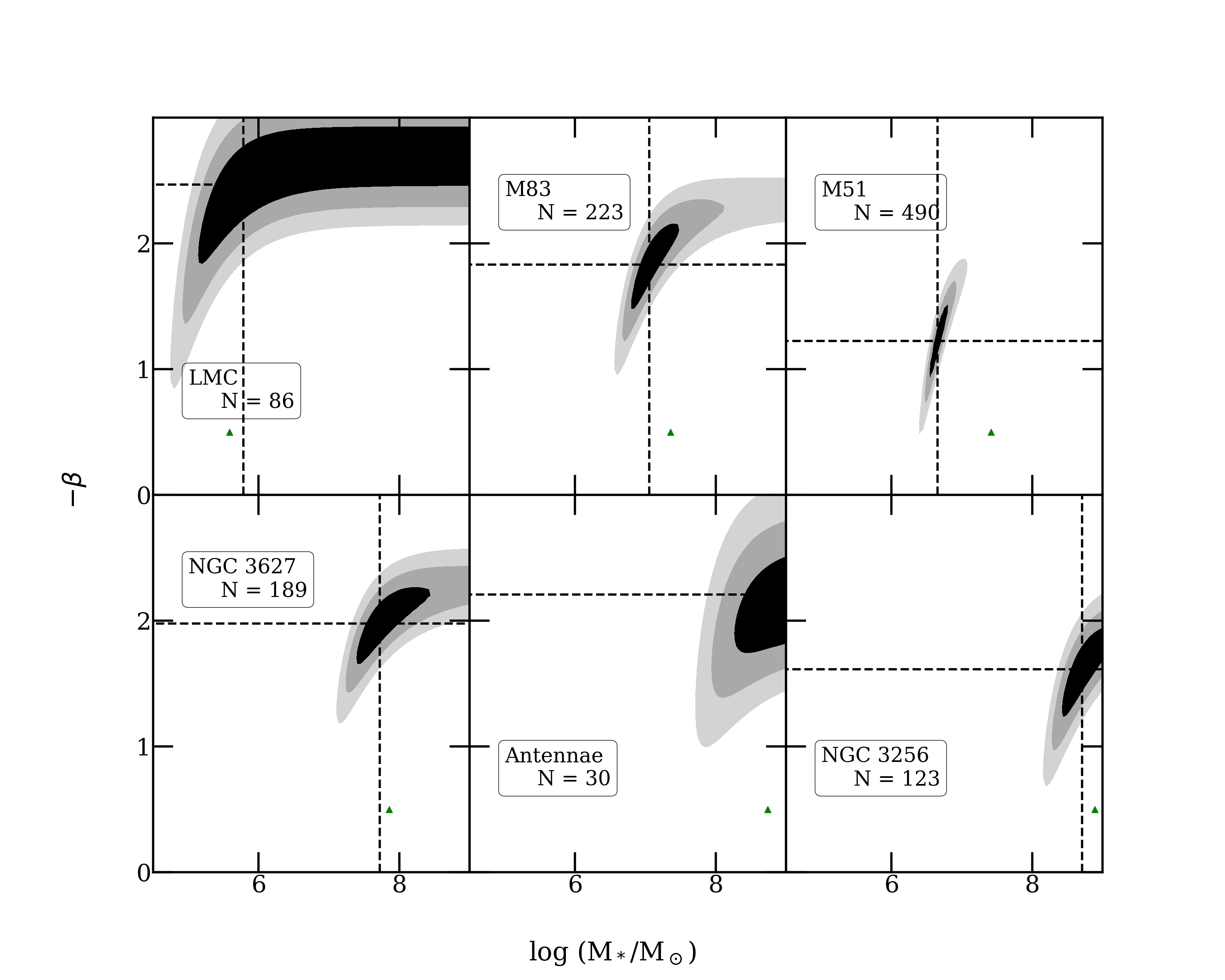}
	\\ Clusters: \\
	\includegraphics[width = 12.5 cm]{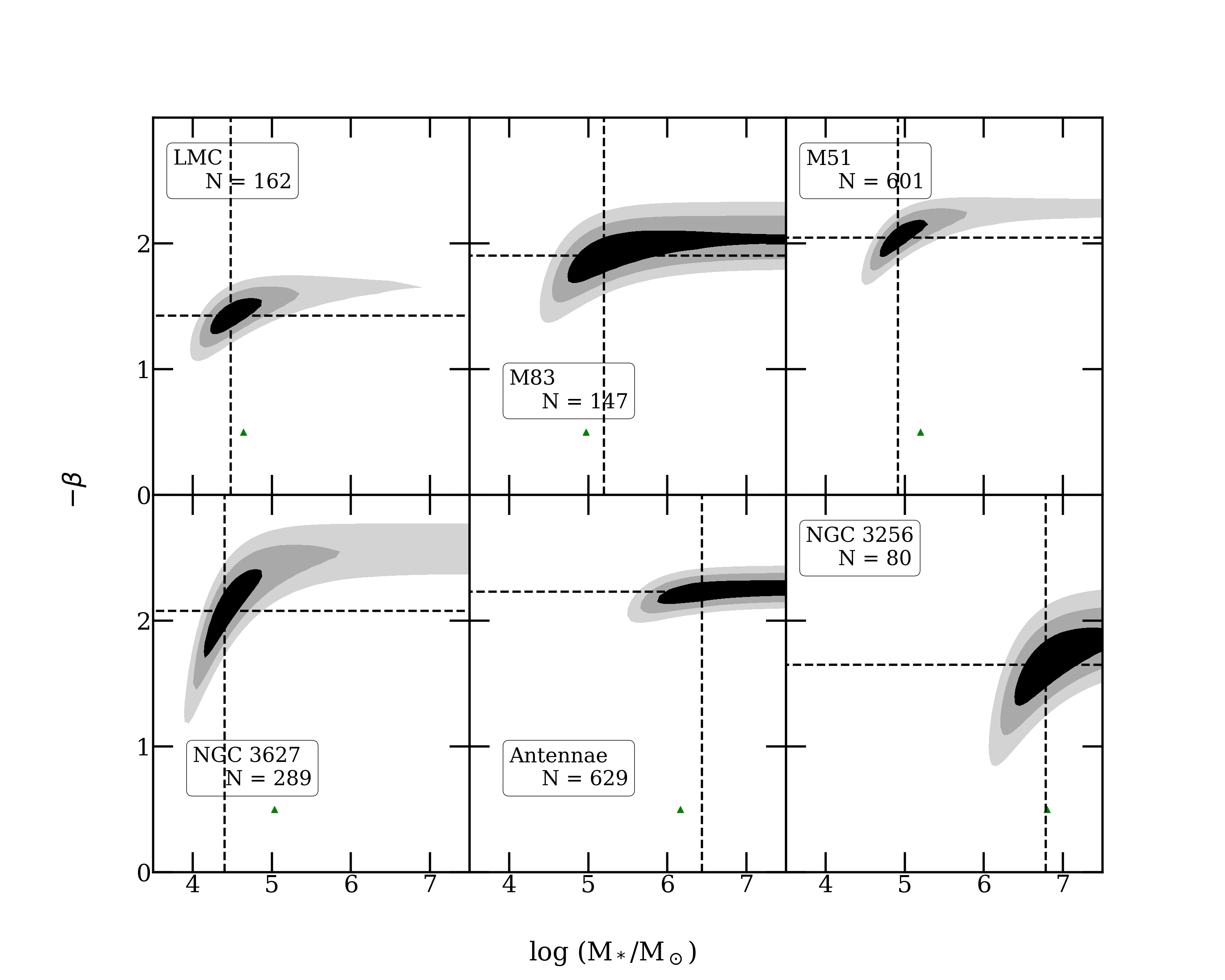}
	\caption{{\bf Top:} The top panels show the results of our maximum-likelihood fits for the Schechter parameters $\beta$ and $M_*$ to the GMC populations of our sample galaxies, and the bottom panels show the results for the young ($\tau < 10$~Myr) cluster populations. No measurement uncertainties are included in the fitting. The dashed lines show the best-fit values of $\beta$ and $M_*$, while the boundaries of the shaded regions show the 1, 2, and 3$\sigma$ confidence contours. The small triangles indicate the most massive GMC or cluster in each catalog.}\label{fig-likelihood}
\end{figure}

Figure~\ref{fig-likelihood} shows the best-fit values of $\beta$ and $M_*$ (dashed lines) for the GMCs (top panels) and clusters (bottom panels), when no uncertainties in the measurements are included. The shaded regions show the 1, 2, and 3$\sigma$ contours resulting from our maximum-likelihood fit. The best-fit values of $\beta$ and $M_*$ and their 1$\sigma$ uncertainties are also listed in Table~\ref{tab-likefitgmc} for GMCs and in Table~\ref{tab-likefitsc} for clusters. Most of the contours have a diagonal portion, which indicates the trade-off between a steeper value of $\beta$ and a higher cutoff mass $M_*$ and vice-versa. There is also a relatively flat portion for some galaxies at higher values of $M_*$, i.e. as the mass function approaches a pure power-law. 
\par
Our results for the GMCs (shown in the top set of panels in Figure~\ref{fig-likelihood}) can be broadly classified into three groups. For the LMC, the Antennae, and NGC~3256, the $1\sigma$ contours (darkest region) remain open up to the right edge of the diagram, i.e. up to the maximum tested value of $M_*$. This means that the value of $M_*$ is indeterminant and the upper portion of the mass function is consistent with a pure power law. We note that the large allowable range in $M_*$, particularly for the LMC and the Antennae, is at least partly driven by the relatively small number of GMCs in those catalogs. For M83 and NGC~3627, the $3\sigma$ and $2\sigma$ contours respectively, remain open up to the maximum tested value for $M_*$. In these cases, the GMC masses are also consistent with being drawn from a pure power law (not excluded at $>3\sigma$ significance), but with some weak evidence for a physical cutoff. M51 is the only galaxy in our sample that shows evidence for an upper cutoff in the mass function of its GMCs at greater than $3\sigma$ significance.
\par
We present the results for clusters younger than 10 Myr in the bottom set of six panels in Figure~\ref{fig-likelihood}. We previously presented results from our maximum-likelihood fitting for the cluster populations in all galaxies (except for NGC 3627) in three intervals of age: $<10$, $10-100$, and $100-400$~Myr \citep{Mok19}. As discussed in that work, ideally we should find consistent results for the Schechter parameters $\beta$ and $M_*$ between all three age intervals, because we do not expect the physics of cluster formation to change significantly over such a short time period (only $\sim3$\% of the Hubble age). However, systematic errors can affect the mass estimates differently in the different age intervals.
\par
For M83, the Antennae, and NGC 3256, the $1\sigma$ contours remain open up to the right edge of the diagram. M51 and NGC 3627 are consistent with a power law, but also show some weak evidence for a cutoff. The only galaxy with $>3\sigma$ evidence for an exponential cutoff $M_*$ in its young cluster population is the LMC, but no cutoff is found in the older populations \citep{Mok19}. {\em We conclude that physical cutoffs appear to be the exception rather than the rule, in the mass functions of GMCs and young clusters in our sample.}
\begin{deluxetable*}{ccccccccc}[ht]
	\tablecaption{GMC Mass Function Parameters, with $\sigma(\log M) = 0.0$ \label{tab-likefitgmc}}
	\tablewidth{0pt}
	\tablehead{
		\colhead{Galaxy} & \colhead{log $M_{\rm lim}$} & \colhead{log $M_{\rm max}$} & \colhead{Num} & \colhead{$-\beta$} & \colhead{log $M_*$} & \colhead{$-\beta_{\rm PL}$}
	}
	\startdata
	LMC & 4.5 & 5.59 & 86 & 2.47 [1.85, 2.95] & 5.78 [5.15, 9.00] & 2.68 [2.50, 2.85] \\
	M83 & 6.1 & 7.36 & 223 & 1.83 [1.45, 2.15] & 7.05 [6.80, 7.50] & 2.35 [2.25, 2.45] \\
	M51 & 6.1 & 7.41 & 490 & 1.22 [0.90, 1.50] & 6.66 [6.55, 6.80] & 2.33 [2.25, 2.40] \\
	NGC 3627 & 6.5 & 7.85 & 189 & 1.98 [1.65, 2.25] & 7.72 [7.40, 8.45] & 2.30 [2.20, 2.40] \\
	Antennae* & 7.0 & 8.74 & 30 & 2.20 [1.70, 2.55] & 9.00 [8.25, 9.00] & 2.23 [2.00, 2.45] \\
	NGC 3256 & 7.6 & 8.89 & 123 & 1.61 [1.25, 1.95] & 8.71 [8.40, 9.00] & 2.10 [2.00, 2.20] \\
	\enddata
	\tablenotetext{}{Square brackets indicate 1$\sigma$ confidence intervals}
	\tablenotetext{*}{Indicates that the Nelder-Mead routine converged to a value beyond the adopted grid. The maximum-likelihood value found in the grid is adopted instead.}
\end{deluxetable*}
\begin{deluxetable*}{ccccccc}[ht]
	\tablecaption{Cluster Mass Function Parameters, with $\sigma(\log M) = 0.0$ \label{tab-likefitsc}}
	\tablewidth{0pt}
	\tablehead{
		\colhead{Galaxy} & \colhead{log $M_{\rm lim}$} & \colhead{log $M_{\rm max}$} & \colhead{Num} & \colhead{$-\beta$} & \colhead{log $M_*$} & \colhead{$-\beta_{\rm PL}$}
	}
	\startdata
	LMC & 2.5 & 4.64 & 162 & 1.42 [1.25, 1.55] & 4.48 [4.20, 4.90] & 1.67 [1.60, 1.70] \\
	M83 & 3.3 & 4.97 & 147 & 1.90 [1.70, 2.10] & 5.20 [4.75, 7.50] & 2.04 [1.95, 2.10] \\
	M51 & 3.5 & 5.20 & 601 & 2.04 [1.90, 2.20] & 4.92 [4.70, 5.30] & 2.28 [2.25, 2.35] \\
	NGC 3627 & 3.5 & 5.04 & 289 & 2.08 [1.70, 2.40] & 4.41 [4.15, 4.90] & 2.56 [2.50, 2.65] \\
	Antennae & 4.0 & 6.17 & 629 & 2.23 [2.15, 2.30] & 6.44 [5.85, 7.50] & 2.26 [2.20, 2.30] \\
	NGC~3256 & 5.2 & 6.80 & 80 & 1.65 [1.30, 1.95] & 6.78 [6.40, 7.50] & 1.93 [1.85, 2.05] \\
	\enddata
	\tablenotetext{}{Square brackets indicate 1$\sigma$ confidence intervals}
\end{deluxetable*}
\subsection{Results With Measurement Uncertainties}\label{subsec-maxlikelihood-result-error}
\par
Thus far, we have neglected uncertainties in the mass estimates of clusters and GMCs. We assess the impact that observational uncertainties have on our results for $\beta$ and $M_*$ by repeating our maximum-likelihood analysis after convolving the Schechter function with a log-normal error distribution of width 0.3 in log~$M$. This is the typical uncertainty found in mass estimates of GMCs in M51 (see Section~\ref{subsec-gmc-catalog}), and for stellar clusters (see Section~\ref{subsec-clusters-catalog}). We follow the procedure for including measurement uncertainties outlined in \citet{Efstathiou88}, and described in detail in Appendix~\ref{sec-appendix}. The main effect of including uncertainties using this method is to modify the shape of the Schechter function such that it more closely resembles a power law, with the exponential cutoff shifting to a higher mass (see Figure~\ref{fig-appendix_1}). This procedure, however, leaves a power-law mass function invariant, including the index $\beta_{\rm PL}$.
\par
\begin{figure}[ht]
	\centering
	GMCs: \\
	\includegraphics[width = 12.5 cm]{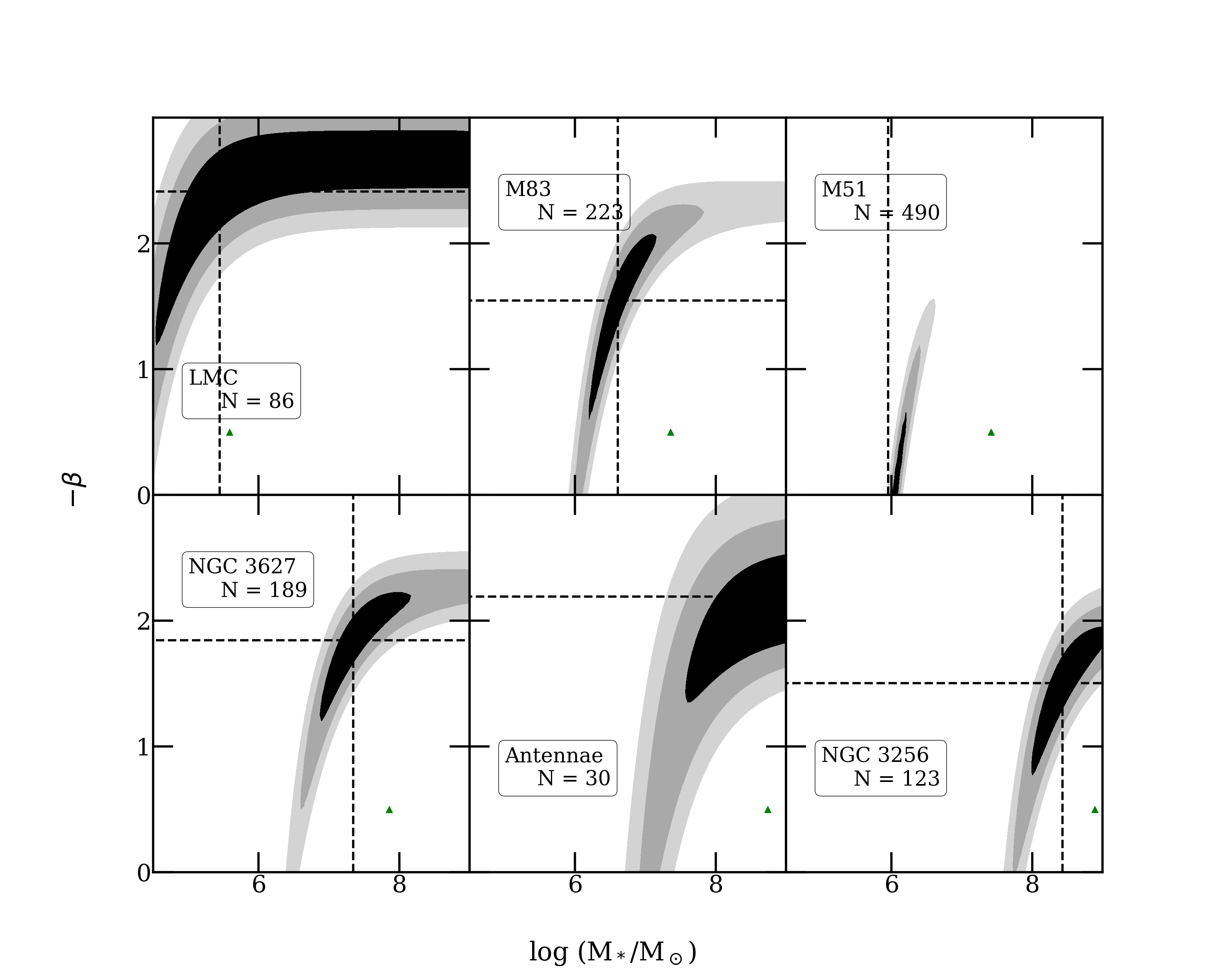}
	\\ Clusters: \\
	\includegraphics[width = 12.5 cm]{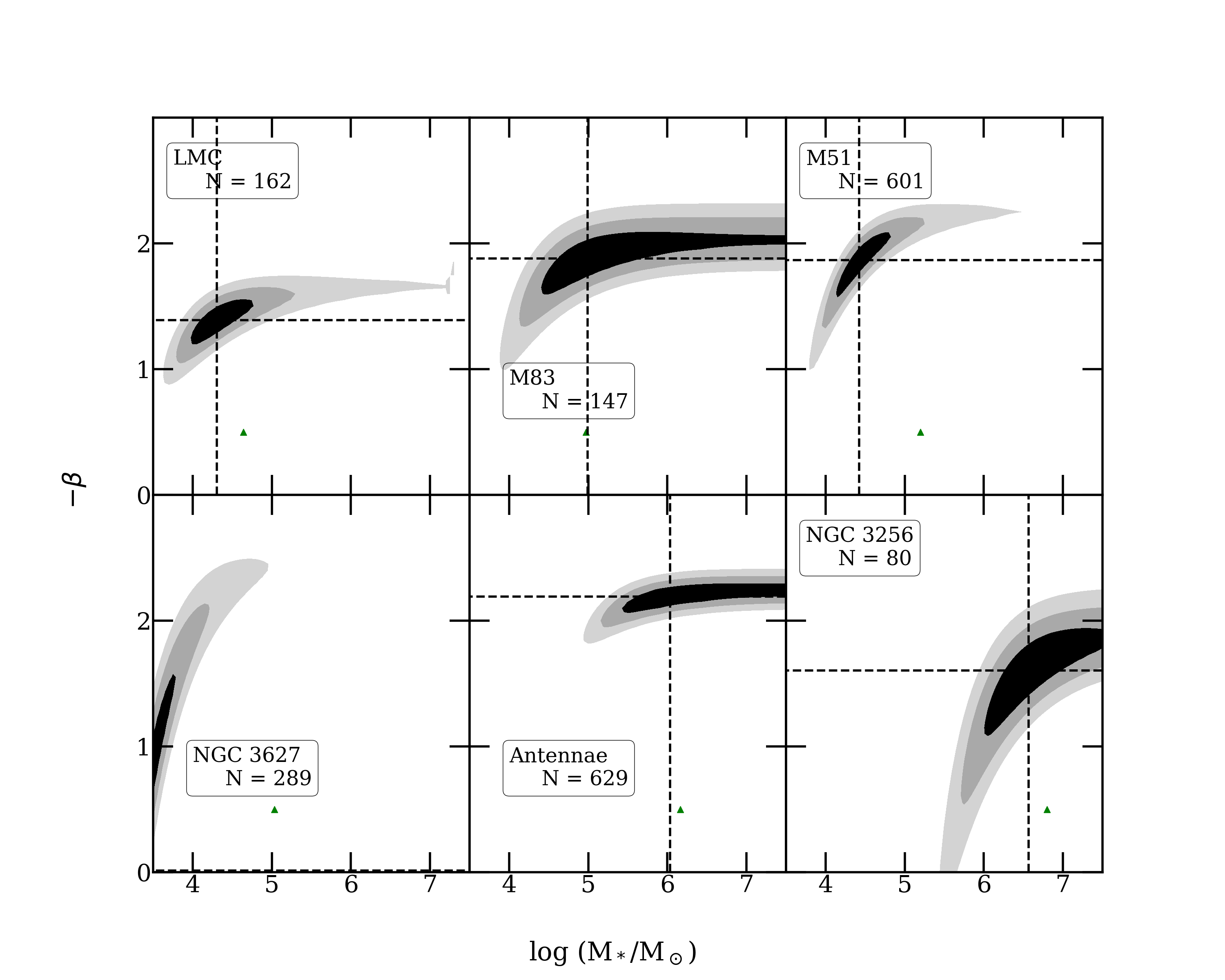}
	\caption{Same as Figure~6, but we fit with a Schechter function convolved with a log normal error distribution with dispersion $\sigma (\log M) = 0.3$.}\label{fig-likelihooderr}
\end{figure}

 
In Figure~\ref{fig-likelihooderr}, we show the results of our maximum-likelihood fits when uncertainties of 0.3 in log~$M$ are included. The corresponding best-fit values of $M_*$ and $\beta_{\rm PL}$ are listed in Table~\ref{tab-likefitgmcerr} for GMCs and in Table~\ref{tab-likefitscerr} for clusters. Compared with the zero-error case, the confidence contours increase in size, allowing for a larger range and hence correspondingly weaker constraints on the power-law index $\beta$ and cutoff mass $M_*$. The GMC mass functions in our sample still do not show evidence for $M_*$, with the exception of M51. In this case, including uncertainties in the fit still results in a detection of $M_*$ at the $>3\sigma$ level, but with a lower best-fit value for $M_*$.
\par
For the cluster populations, a similar behavior occurs, where including uncertainties lowers the best-fit values of $M_*$ in M51 and NGC~3627. This happens because a Schechter function convolved with the error distribution has a shape closer to that a power law, which shifts the best-fit $M_*$ to a lower value in order to compensate for this behavior. While accounting for observational uncertainties using our simple prescription results in a detection of $M_*$ in the {\em cluster} population of NGC 3627, no corresponding detection is made for GMCs in this galaxy. {\em M51 remains the only galaxy in our sample with evidence for a high-mass cutoff in both the GMC and young cluster populations.}
\begin{deluxetable*}{ccccccc}[ht]
	\tablecaption{GMC Mass Function Parameters, with $\sigma(\log M) = 0.3$ \label{tab-likefitgmcerr}}
	\tablewidth{0pt}
	\tablehead{
		\colhead{Galaxy} & \colhead{log $M_{\rm lim}$} & \colhead{log $M_{\rm max}$} & \colhead{Num} & \colhead{$-\beta$} & \colhead{log $M_*$}
	}
	\startdata
	LMC & 4.5 & 5.59 & 86 & 2.41 [1.20, 2.90] & 5.44 [4.50, 9.00] \\
	M83 & 6.1 & 7.36 & 223 & 1.55 [0.60, 2.10] & 6.61 [6.20, 7.15] \\
	M51* & 6.1 & 7.42 & 490 & 0.00 [0.00, 0.75] & 5.95 [6.00, 6.25] \\
	NGC 3627 & 6.5 & 7.85 & 189 & 1.84 [1.20, 2.25] & 7.34 [6.85, 8.15] \\
	Antennae* & 7.0 & 8.74 & 30 & 2.20 [1.35, 2.55] & 9.00 [7.55, 9.00] \\
	NGC 3256 & 7.6 & 8.89 & 123 & 1.50 [0.75, 1.95] & 8.43 [8.00, 9.00] \\
	\enddata
	\tablenotetext{}{Square brackets indicate 1$\sigma$ confidence intervals}
	\tablenotetext{*}{Indicates that the Nelder-Mead routine converged to a value beyond the adopted grid. The maximum-likelihood value found in the grid is adopted instead.}
\end{deluxetable*}
\begin{deluxetable*}{ccccccc}[ht]
	\tablecaption{Cluster Mass Function Parameters, with $\sigma(\log M) = 0.3$ \label{tab-likefitscerr}}
	\tablewidth{0pt}
	\tablehead{
		\colhead{Galaxy} & \colhead{log $M_{\rm lim}$} & \colhead{log $M_{\rm max}$} & \colhead{Num} & \colhead{$-\beta$} & \colhead{log $M_*$}
	}
	\startdata
	LMC & 2.5 & 4.64 & 162 & 1.39 [1.20, 1.55] & 4.30 [3.95, 4.75] \\
	M83 & 3.3 & 4.97 & 147 & 1.88 [1.60, 2.10] & 4.99 [4.40, 7.50] \\
	M51 & 3.5 & 5.20 & 601 & 1.87 [1.55, 2.10] & 4.43 [4.10, 4.85] \\
	NGC 3627* & 3.5 & 5.04 & 289 & 0.00 [0.00, 1.50] & 3.50 [3.50, 3.80] \\
	Antennae & 4.0 & 6.17 & 629 & 2.19 [2.05, 2.30] & 6.04 [5.40, 7.50] \\
	NGC~3256 & 5.2 & 6.80 & 80 & 1.60 [1.10, 1.95] & 6.56 [6.00, 7.50] \\
	\enddata
	\tablenotetext{}{Square brackets indicate 1$\sigma$ confidence intervals}
	\tablenotetext{*}{Indicates that the Nelder-Mead routine converged to a value beyond the adopted grid. The maximum-likelihood value found in the grid is adopted instead.}
\end{deluxetable*}
\subsection{Schechter Function vs. Power-law Fits}\label{subsec-maxlikelihood-powerlaw}
\par
Since we do not find a physical cutoff for most of the galaxies in our sample, we also apply our maximum-likelihood fitting routine for the case where $M_* \rightarrow \infty$, i.e. fitting a pure power-law to the data. We list the best-fit power-law index $\beta_{\rm PL}$ in the last column of Tables~\ref{tab-likefitgmc} and \ref{tab-likefitsc}. In every case, the fitting returns a steeper value for $\beta_{\rm PL}$ compared with the best fit index $\beta$ for the Schechter case. This is due to the correlation between $\beta$ and $M_*$, which leads to flatter values of $\beta$ when an underlying Schechter function is assumed. Therefore, caution should be used when comparing results from different works, since adopting different underlying functions (pure power law vs. Schechter vs truncated power law) can lead to different values of the power-law index. 
\begin{table}[ht]
	\caption{AIC and Relative Likelihood Values}\label{tab-AIC}
	\centering
	\begin{tabular}{lccc}
	 \hline\hline
		Clusters & AIC$_{\rm PL}$ & AIC$_{\rm Sch}$ & Rel. \\
		\hline
		LMC & 2801 & {\bf 2791} & $6\times10^{-3}$ \\
		M83 & 2806 & {\bf 2805} & $9\times10^{-1}$ \\
		M51 & 11540 & {\bf 11532} & $2\times10^{-2}$ \\
		NGC 3627 & 5353 & {\bf 5348} & $8\times10^{-2}$ \\
		Antennae & {\bf 13549} & 13550 & $6\times10^{-1}$ \\
		NGC 3256 & {\bf 2272} & 2270 & $5\times10^{-1}$ \\
		\hline
	\end{tabular}
	\begin{tabular}{lccc}
	 \hline\hline
		GMCs & AIC$_{\rm PL}$ & AIC$_{\rm Sch}$ & Rel. \\
		\hline
		LMC & \bf{1973} & 1975 & $5\times10^{-1}$ \\
		M83 & 6913 & \bf{6906} & $4\times10^{-2}$ \\
		M51 & 15211 & \bf{15165} & $1\times10^{-10}$ \\
		NGC 3627 & 6230 & {\bf 6227} & $3\times10^{-1}$ \\
		Antennae & {\bf 1069} & 1071 & $3\times10^{-1}$ \\
		NGC 3256 & 4754 & {\bf 4749} & $8\times10^{-2}$ \\
		\hline
 \end{tabular}
 \\ \vspace{0.25cm} Note: Bolded values indicate the lower value of the AIC. The Relative Likelihood test shows the \\
 probability that the less likely model also minimizes information loss.
\end{table}
\par
We perform statistical tests to assess if one of the underlying distributions (power law or Schechter) gives a better description of the data. We calculate the Akaike Information Criterion (AIC; \citet{Akaike74}), given by ${\rm AIC} = 2k - 2\ln L_{\rm max}$, where $k$ is the number of parameters and $L_{\rm max}$ is the maximum value of the likelihood function. A lower AIC value indicates a better description of the data. The results for young ($\tau < 10$~Myr) clusters and GMCs are compiled in Table~\ref{tab-AIC} for the case without measurement uncertainties. We also compile results from the relative likelihood test \citep{Burnham02}, given by: $\exp((\rm{AIC}_1 - \rm{AIC}_2)/ 2)$, which is the probability that the less likely model {\it also} minimizes the information and provides a good description of the data. Lower values from the relative likelihood test indicate a stronger preference for the model with a lower AIC value. We adopt a likelihood value of 0.05 or less (corresponding to $\approx2\sigma$) as suggesting a preference for one model over the other. For clusters, the results indicate a weak preference for a Schechter function in the LMC and M51. For GMCs, the results suggest a strong preference for an underlying Schechter function in M51 and a weak preference in M83. These results largely agree with our qualitative discussion above of the maximum-likelihood fits.
\par
For GMCs, when a Schechter function is assumed to describe the mass function, the best-fit $\beta$ values in our galaxy sample range from $-1.22$ to $-2.47$ with a median $\beta \approx -2.2$. When we use a power-law instead, the best-fit $\beta_{\rm PL}$ values range from $-1.93$ to $-2.68$ with a median $\beta_{\rm PL} \approx -2.3$ and a standard deviation of $\approx 0.3$. Thus for our galaxy sample, we find $\beta_{\rm PL}=-2.3\pm0.3$ for GMC mass functions. 
\par
For young clusters, we find best-fit $\beta$ values (with an underlying Schechter function) that range from $-1.42$ to $-2.23$ with a median $\beta\approx -1.9$, and $\beta_{\rm PL}$ values (power law) from $-1.67$ to $-2.56$ with a median $\beta_{\rm PL}\approx -2.0$ and standard deviation $\approx0.3$. Therefore, we find $\beta_{\rm PL}=-2.0\pm0.3$ for the cluster mass functions. \footnote{In our previous papers characterizing cluster systems, we found uncertainties closer to $\sim\pm0.2$ for $\beta_{\rm PL}$. Here, we find a somewhat larger uncertainty because of the smaller number of measurements, i.e. we determine $\beta_{\rm PL}$ for a single age range ($\tau < 10$ Myr) rather than all age ranges for each galaxy.}
\begin{figure*}[ht]
	\centering
	\includegraphics[width=7.5cm]{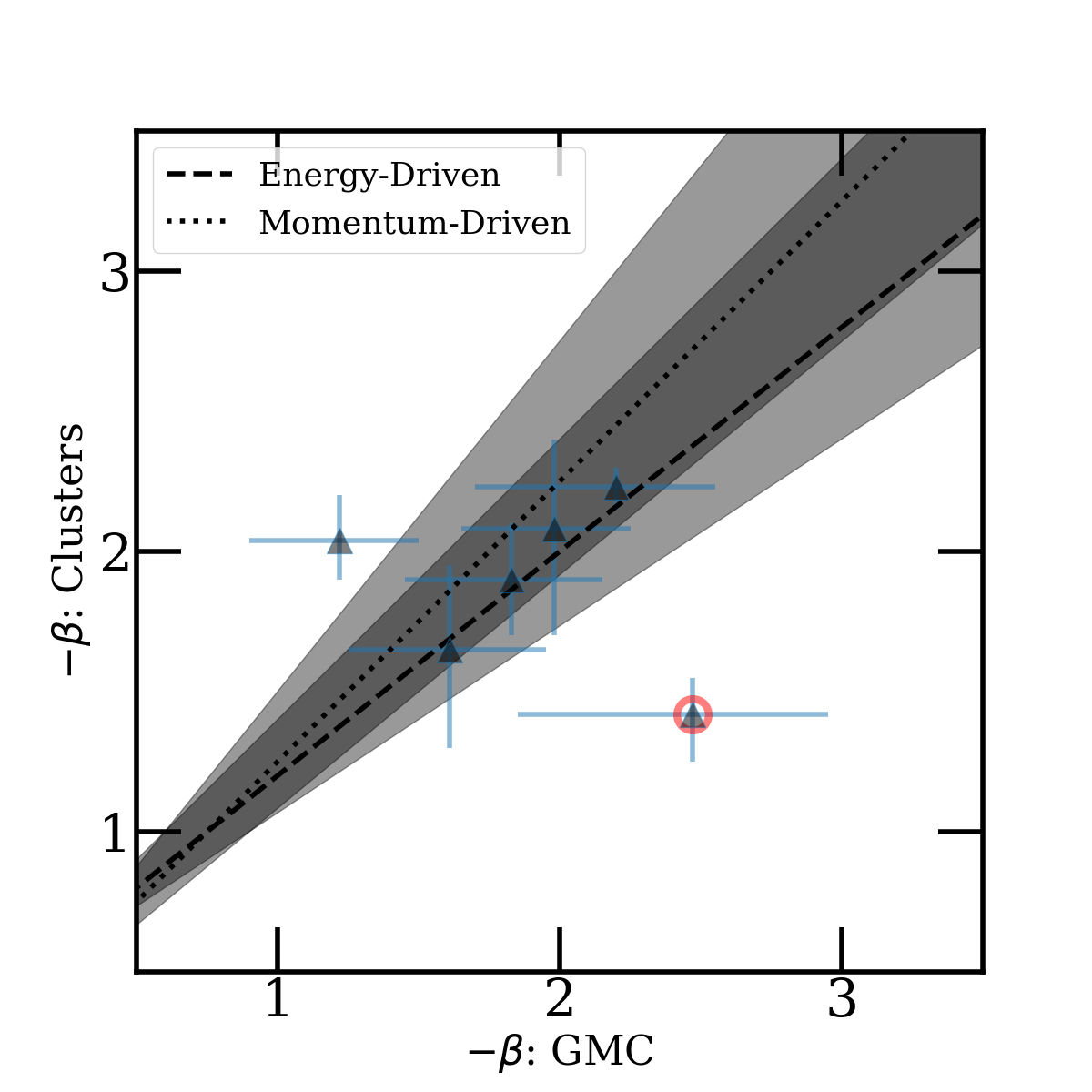}
	\includegraphics[width=7.5cm]{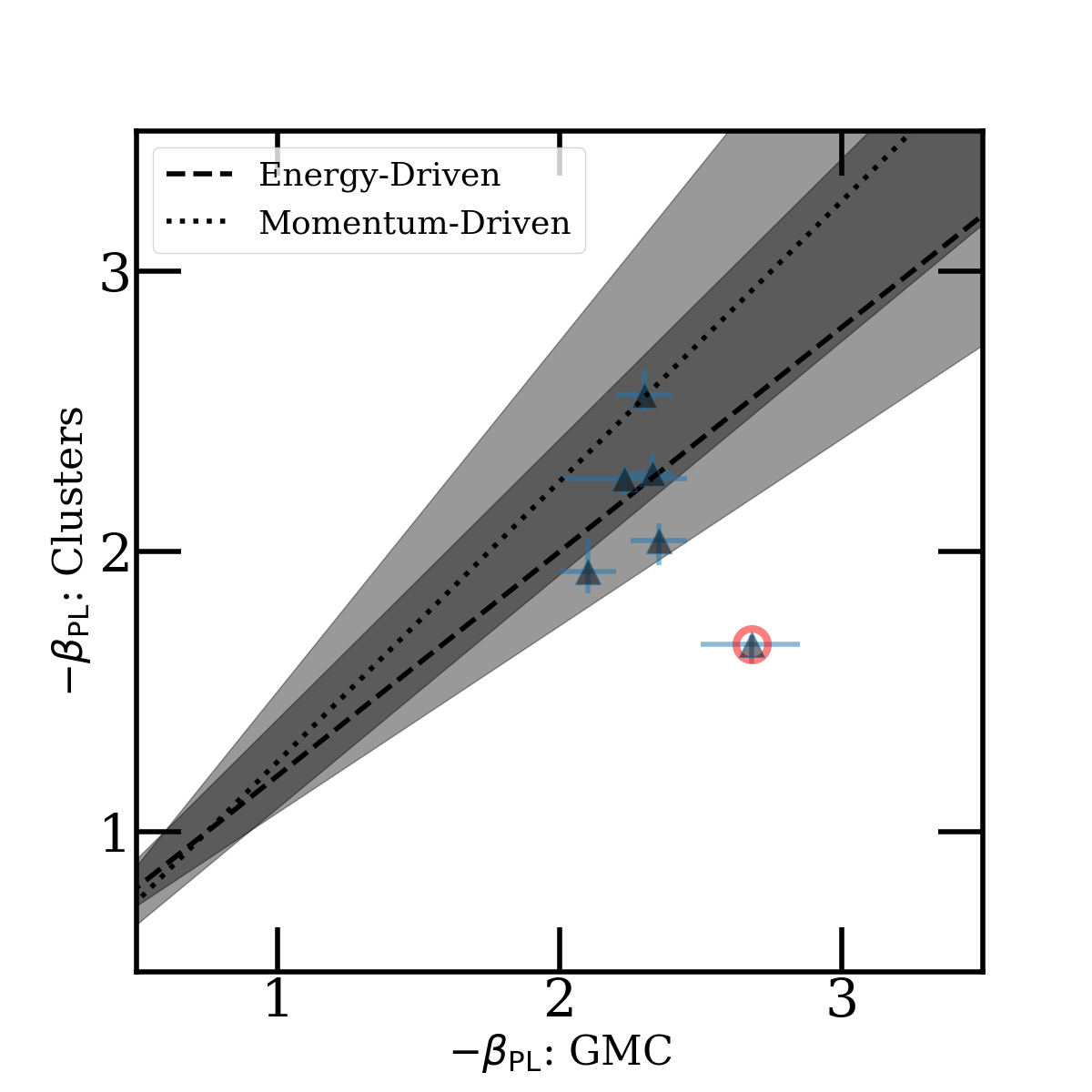}
	\caption{A comparison of the best-fit power-law indices (-$\beta$) in the mass functions of GMCs and young ($\tau < 10$~Myr) clusters. The left panel shows the results when we assume an underlying Schechter function, and the right panel shows the results when an underlying power law is assumed. The results for the LMC are circled in red, since this galaxy has more piecemeal coverage and significantly higher resolution than the others, as described in the text. The dashed and dotted lines show predictions for the relation between $\beta_{\rm C}$ and $\beta_{\rm GMC}$ in the energy-driven and momemtum-driven feedback regimes, where the exponent of the index of the mass-radius relation of the protoclusters is assumed to be $\alpha = 0.5$ (see Section~\ref{subsec-discussion-gmcvscluster}). The shaded regions show the predictions when uncertainties of $\pm0.1$ are allowed on $\alpha$.}\label{fig-sum1}
\end{figure*}
\section{Discussion}\label{sec-discussion}
\subsection{The Shape of the GMC Mass Functions}\label{subsec-discussion-gmcmfshapes}
\par
Earlier studies have found a large range for the power law index of GMC mass functions in nearby galaxies, from $\beta_{\rm PL}=-1.5$ in the inner disk of the Milky Way \citep{Rice16} to $\beta_{\rm PL}=-2.6$ in the LMC \citep{Wong11} and $\beta_{\rm PL}=-2.9$ in M33 \citep{Rosolowsky05}. There are also some discrepancies among the published indices, such as $\beta_{\rm PL} = -1.7$ for the LMC \citep{Blitz07} and $\beta_{\rm PL} = -2.3$ for M33 \citep{Gratier12}. Some of these discrepancies may be caused by different underlying assumptions, observational techniques, and fitting methods. Thus, from earlier work it is unclear if the mass functions of GMCs have similar or different shapes from one galaxy to another.
\par 
In this work, we have applied a uniform methodology to fit the mass functions of GMCs and young clusters in six star-forming galaxies (LMC, M83, M51, NGC~3627, the Antennae, and NGC~3256), which includes irregulars, spirals, mergers, dwarf, and massive galaxies. {\em One of our main results is that a pure power law provides a good fit to the mass functions of GMCs for all but one galaxy (M51) in our sample.} We also find a relatively small range for the power-law index, $\beta_{\rm PL}=-2.3\pm0.3$, indicating that the shapes of GMC mass functions are fairly similar among the galaxies in our sample.
\par 
Overall, we find good agreement when we compare our fitting results for GMC catalogs in the LMC, M83, and M51 with previously published ones. For the LMC and M51, previous works have fitted a truncated power law of the form $N(M'>M) = N_o \bigg[(M / M_o)^{\beta + 1} -1 \bigg]$ \citep{Rosolowsky05}, which returns a truncation mass $M_o$ and the statistic $N_o$ (values of $N_o$ significantly larger than 1 indicate that a cutoff is preferred over a power law). For the LMC, \citet{Wong13} found $N_o = 0.15 \pm 1.48$ using the same catalog used in this paper, i.e. no cutoff was detected, similar to our result. They also found $\beta = -2.57 \pm 0.20$, similar to our best fits of $\beta=-2.47$ and $\beta_{\rm PL}=-2.68$. For M51, \citet{Colombo14} found $N_o = 17 \pm 7$ ($\approx2.5\sigma$ significance), $M_o = (18.5 \pm 3.4) \times 10^6~M_\odot$, and $\beta = -2.29 \pm 0.09$. The maximum-likelihood method used here also supports a truncation in the GMC mass function in M51, but with a somewhat lower cutoff value ($M_*=4.6\times 10^{6}$ vs. $M_o = 1.9\times10^7$), possibly due to the different assumed functional form. For M83, \citet{Freeman17} did not present fit results for the entire GMC sample, instead focusing on fits in six radial bins, but their results appear to be consistent with our weak indication for a physical cutoff.
\subsection{GMCs vs. Clusters}\label{subsec-discussion-gmcvscluster}
\par
A comparison between the shapes of GMCs and clusters in the same galaxies provides important clues to the processes that operate during cluster formation and early evolution. If the shapes of the mass functions of GMCs and clusters are similar, this implies that the SFE has little or no dependence on the mass of the protoclusters. A comparison of the normalizations between the GMC and cluster mass functions (in the same galaxy) would then provide a numerical estimate of the SFE. Unfortunately, the available datasets are too heterogeneous, and include molecular gas observations with different CO transitions, different angular and physical resolutions, and different sky coverage of the GMC and cluster maps, which all introduce significant uncertainties in the normalization of the GMC mass function. Thus, we do not attempt to estimate a numerical values for the star formation efficiencies in this paper.
\par
As clusters form, the energy and momentum injected by young stars eventually expel the remaining gas and halts further star formation. Several feedback mechanisms (e.g., protostellar jets and outflows, radiation pressure, photoionized gas, and supernovae) likely operate simultaneously in combination during these early phases. Radiative losses inside the young clusters determine how much of the energy from stellar feedback is available to remove the gas. The FKM model describes two extreme regimes that likely encompass most realistic situations: the energy-driven regime, where there are no radiative losses so all of the feedback energy stays in the region and is available to expel the ISM, and the momentum-driven regime, where there are maximum radiative losses. In the model, the mass-radius relation of the protoclusters is approximated by a power-law, $R_h \propto M^{\alpha}$. Studies in the Milky Way and LMC of star-forming clumps \citep[e.g. FKM; ][]{Wu10, Wong19} and GMCs \citep[]{Larson81, Blitz07} indicate $\alpha\approx0.5$. FKM derived the following relations between the power-law indices of the mass functions of the gas-dominated protoclusters (G) and the resulting stellar clusters (S) for the energy-driven case:
\begin{equation}
 \beta_{\rm S} = \frac{2(\beta_{\rm G} + \alpha - 1)}{5(1 - \alpha)},
\end{equation}
and the following relation for the momentum-driven case: 
\begin{equation}
 \beta_{\rm S} = \frac{2\beta_{\rm G} + \alpha - 1}{4(1 - \alpha)}.
\end{equation}
\par
Strictly speaking, the gas-dominated protoclusters in the model correspond most directly to dense, star-forming clumps within GMCs rather than to the GMCs themselves. However, the clump mass function can also be fitted by a power-law function with $\beta_{\rm PL} \approx 1.7$ \citep[e.g.][]{Wong08}. Since GMCs and clumps have similar mass functions and mass-radius relations, it is likely that the scaling relations for the power law indices of the mass functions in Equations (3) and (4) derived in the simple feedback model apply at least approximately to both types of clouds. Therefore, in Figure~\ref{fig-sum1}, we plot the predicted FKM relations on top of the comparison of $\beta_{\rm S}$ and $\beta_{\rm G}$ in our sample.
\par
We also note that in the FKM model, the SFE ($\epsilon$) in protoclusters depends primary on their mean surface density ($\Sigma$). One important consequence of this relation is that if $\Sigma$ is roughly constant from one protocluster to another, then $\epsilon$ will also be roughly constant and the power law indices of the mass functions of molecular clouds and young star clusters will be similar (the case of $\alpha=0.5$ discussed above). Recent observations of clouds in nearby galaxies, such as the sample of 15 galaxies compiled in \citet{Sun18} from the PHANGS survey and archival sources, show large variations in $\Sigma$ among galaxies, ranging from $\sim10^1$ to $\sim10^4$ $M_\odot$ pc$^{-2}$. However, if we restrict attention to the clouds within any single galaxy (excluding clouds near the center), then the distribution of $\Sigma$ is much narrower, consistent with the FKM model; see Figure 1 and 2 in \citet{Sun18}. Given that we apply the FKM model to each galaxy on an individual basis, this framework is valid for the analysis presented here.
\par
We see that the momentum-driven regime (dotted line) predicts a somewhat steeper relation between the power-law indices of clusters and GMCs (in the sense that the clusters are predicted to have steeper distributions) than the energy-driven regime (dashed line). We also show uncertainties of $\pm0.1$ on $\alpha$ as the shaded regions. We plot our best-fit results for $\beta$ (with an underlying Schechter function) on the left and for $\beta_{\rm PL}$ (with an underlying power-law) on the right in Figure~\ref{fig-sum1}. The LMC is an outlier in this plot, possibly due to the partial coverage and significantly higher physical resolutions. Our results for the other five galaxies are largely consistent with either set of predictions, but the sample is fairly small. We find tentative signs that the mass functions of GMCs may be slightly steeper than those of clusters in the same galaxies, with $\beta_{\rm PL}=-2.3\pm0.3$ for GMCs and $\beta_{\rm PL}=-2.0\pm0.3$ for clusters, but overall we find that $\beta_{\rm Clusters} \approx \beta_{\rm GMC} \approx-2$. The similarity in the shapes of the mass functions of GMCs and young clusters suggests that the SFE for the clusters in our sample is largely independent of the cloud or protocluster mass, consistent with the FKM model.
\subsection{Star Formation Efficiency and $M_*$}\label{subsec-discussion-sfe}
\par
Previous works have estimated the SFE in GMCs from the ratio of the upper cutoffs in cluster and GMC mass functions \citep[e.g.][]{Freeman17,Messa18b}. Note that many of the previous works on this topic have adopted a truncated power law instead of a Schechter function, but the general results are similar (see Section~\ref{subsec-discussion-gmcmfshapes}). We will continue to use the terminology of $M_*$ to denote a steep decline at the high mass end (i.e. a physical cutoff). \citet{Messa18b} found a ratio of $\sim1\%$, which they attributed to a combination of a $\sim10\%$ SFE plus $\sim10\%$ of stars surviving in bound clusters.
\par
Our results suggest, however, that physical upper cutoffs may be the exception rather than the rule. In cases where $M_*$ is indeterminant or only weakly detected, this method for estimating the SFE is unreliable. The method becomes even more problematic when studying parts of galaxies, where smaller samples result in larger uncertainties in the fitted parameters. In the ideal case, where there is strong evidence for physical upper-mass cutoffs in both the GMC and cluster mass functions, the ratio would represent a real relation between the two populations. However, the results from our sample indicate that this may not be the case for most galaxies.
\section{Conclusions and Summary}\label{sec-conclusions}
\par
In this paper, we measure the shapes of the mass functions of GMCs and young clusters (ages $<10$~Myr) in six star-forming galaxies: LMC, M83, M51, NGC~3627, the Antennae, and NGC~3256. These galaxies span a large range in distance (from $\sim50$~kpc to $\sim40$~Mpc), SFR ($\sim0.25$ to $\sim50~M_{\odot}~\mbox{yr}^{-1}$), and morphology (irregulars, spirals, and mergers). For NGC~3627, we present a new GMC catalog based on archival ALMA observations and a new cluster catalog from the $HST$-based LEGUS and H$\alpha$-LEGUS projects. We also present a new catalog of GMCs in the merging NGC~3256 system based on archival ALMA observations. We perform maximum-likelihood fits of the Schechter function ($dN/dM \propto M^{\beta} \mbox{exp}(-M/M_*)$) and a pure power law ($dN/dM \propto M^{\beta_{\rm PL}}$), to the observed GMC and cluster mass functions in a uniform way, using the procedure from \citet{Mok19}. Our main conclusions are as follows:
\begin{itemize}
	\item We find that, to first order, the majority of the GMC mass functions studied here are consistent with having a power law index $\beta_{\rm PL}=-2.3\pm0.3$. The uncertainties on each $\beta_{\rm PL}$ are large enough that there may be real variations among galaxies, although the range found here is not as large as the full range of $\beta_{\rm PL}$ values found in earlier studies. We find that the mass functions of young clusters in our sample can be described by $\beta_{\rm PL}=-2.0\pm0.3$.
	\item For almost all of our target galaxies ($5/6$), we find little or no evidence for a physical upper cutoff in the mass function of GMCs. This suggests that such cutoffs may be the exception, rather than the rule, in populations of GMCs in nearby galaxies. Previously, we found a similar result for young cluster populations in a sample of eight galaxies, i.e. that most of their mass functions do not show evidence for a physical cutoff at the high end $M_*$ \citep{Mok19}. M51 is the only galaxy in our sample that shows evidence for a physical upper-mass cutoff in both the GMC and young cluster populations.
	\item In general, we find $\beta_{\rm GMC} \approx \beta_{\rm C} \approx -2$. This is consistent with predictions of the analytic FKM model. Since the shapes of the two mass functions are fairly similar, this indicates that the SFE is largely independent of prototcluster mass.
	\item Given the lack of strong cutoffs in our sample, the method of estimating the SFE in a galaxy from the ratio of the cutoff mass ($M_*$) in the cluster to that in the GMC populations may not be applicable except in rare cases.
\end{itemize}
\par
Our results are based on the GMC and cluster populations in a relatively small number of galaxies. We plan to confirm these results in larger samples of galaxies when the data become available. Since it is possible that resolution-dependent and other observational biases may affect the observed shape of the distribution, we believe another important step is to collect higher resolution observations of molecular gas, particularly at the clump scale. Observations of very nearby galaxies or simulated galaxies could also be degraded to simulate a range of different distances, in order to test its impact on the shape of the mass function.
\section*{Acknowledgements}
R.C. acknowledges support from NSF grant 1517819. The authors acknowledge the helpful comments provided by the referee, Erik Rosowlowsky.
\par
The Digitized Sky Surveys were produced at the Space Telescope Science Institute under U.S. Government grant NAG W-2166. The images of these surveys are based on photographic data obtained using the Oschin Schmidt Telescope on Palomar Mountain and the UK Schmidt Telescope. The plates were processed into the present compressed digital form with the permission of these institutions. 
\par
This paper makes use of the following ALMA data: ADS/JAO.ALMA\#2015.1.00956.S and ADS/JAO.ALMA\#2015.1.00714.S. ALMA is a partnership of ESO (representing its member states), NSF (USA) and NINS (Japan), together with NRC (Canada), MOST and ASIAA (Taiwan), and KASI (Republic of Korea), in cooperation with the Republic of Chile. The Joint ALMA Observatory is operated by ESO, AUI/NRAO and NAOJ.
\bibliography{master}
\appendix
\section{Maximum-Likelihood Method with Errors}\label{sec-appendix}
\par
To include measurement errors in our maximum-likelihood fits of the Schechter mass function to GMC and cluser catalogs, we follow the treatment in \citet{Efstathiou88}, who fitted Schechter luminosity functions to galaxy catalogs. As in the error-free case, we adopt the Schechter function ($\psi_t$) as the true underlying mass function of GMCs and clusters:
\begin{equation}
	\psi_t(M) = \frac{dN}{dM} = \bigg(\frac{\psi_*}{M_*}\bigg)~\bigg(\frac{M}{M_*}\bigg)^{\beta}~\exp\bigg(-\frac{M}{M_*}\bigg).
\end{equation}
We represent the distribution of measurement errors by $p_e(M|M')$, where $p_e(M|M')~dM$ is the probability that a cluster with a true mass $M'$ has a measured mass in the small interval $M$ to $M + dM$. The predicted observed mass function ($\psi_o$) is then given by the following integral
\begin{equation}
	\psi_{\rm o}(M) = \int_0^\infty p_e(M|M')~\psi_t(M')~dM' .
\end{equation}
For the error-free case, we have $p_e(M|M') = \delta (M-M')$, where $\delta$ denotes the usual Dirac delta function, and hence $\psi_{\rm o}(M) = \psi_{\rm t}(M)$, as expected.
\par
\begin{figure*}[ht]
	\centering
	\includegraphics[width=8.5cm]{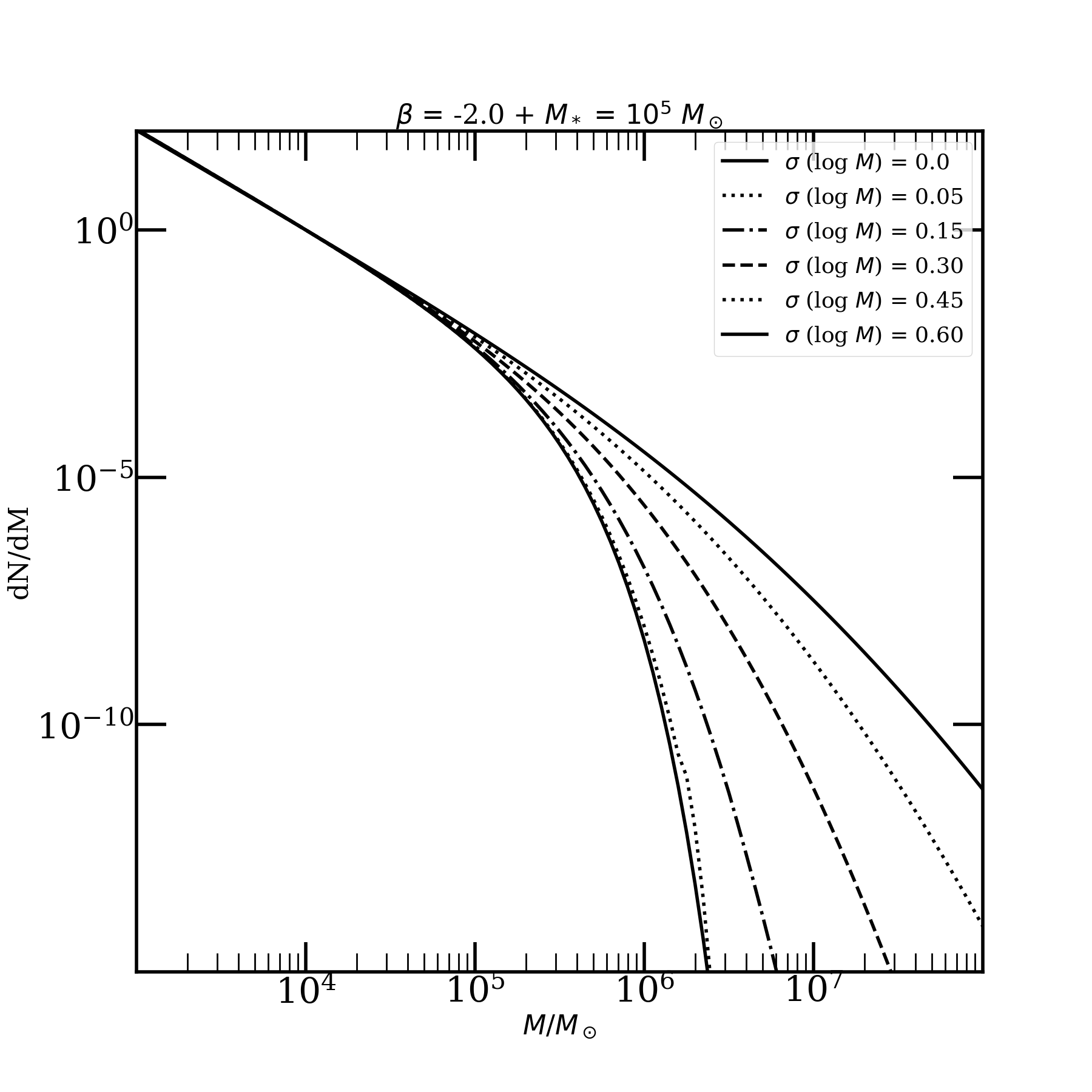}
	\caption{The effects of varying the $\sigma$ parameter on the shape of the observed mass function ($\psi_o$) with $\beta = -2$ and $M_* = 10^5~M_\odot$, where the no-error case is shown as the thick, solid line. Increasing the uncertainty increases the cutoff mass, making the curve appear more and more like a pure power law. }\label{fig-appendix_1}
\end{figure*}

The primary error distribution that we adopt is the log-normal distribution:
\begin{equation}
	p_e(M|M') = \frac{1}{\sqrt{2\pi}\sigma_e M}~\exp\bigg[-\frac{(\ln M - \ln M')^2}{2\sigma_e^2}\bigg]
\end{equation}
Here, $\sigma_e$ is equal to $\sigma\ln 10$, where $\sigma$ is the typical base-10 log uncertainty in the cluster and GMC mass measurements. The predicted observed mass functions from equations (A1) $-$ (A3) is then
\begin{equation}
	\psi_{\rm o}(M) = \frac{1}{\sqrt{2\pi}\sigma_e M}~\bigg(\frac{\psi_*}{M_*}\bigg)~\int_0^\infty \bigg(\frac{M'}{M_*}\bigg)^{\beta}~\exp\bigg(-\frac{M'}{M_*}\bigg)~\exp\bigg[-\frac{(\ln M - \ln M')^2}{2\sigma_e^2}\bigg] dM'.
\end{equation}
In Figure~\ref{fig-appendix_1}, we show the effect that increasing the uncertainty $\sigma$ has on the shape of a Schechter function with $\beta = -2.0$ and $M_* = 10^5~M_\odot$. Larger uncertainties cause the observed mass function ($\psi_o$) to appear more like a power-law, shifting the exponential cutoff to higher masses.
\par
We determine the best-fitting values of $\beta$ and $M_*$ by comparing $\psi_o$ with the data by the same procedure in both the cases with and without measurement errors, namely, by maximizing the likelihood
\begin{equation}
L(\beta,M_*)=\prod_i~P_i,
\end{equation}
with the probability for the $i$th cluster given by
\begin{equation}
P_i = \frac{\psi_o(M_i)}{\int_{M_{\rm min}}^{M_{\rm max}} \psi(M) dM}.
\end{equation}
While we primarily adopt the log-normal distribution for the measurement uncertainties, we also experimented with a few other functional forms, including simple step functions. We find that these alternative error distributions have less effect on the confidence contours for $\beta$ and $M_*$ than the log-normal distribution. Thus, we present in Section~\ref{sec-maxlikelihood} the two bracketing cases, i.e. the no-error case and the log-normal error case.
\par
\begin{figure*}[ht]
	\centering
	\includegraphics[width=16.5cm]{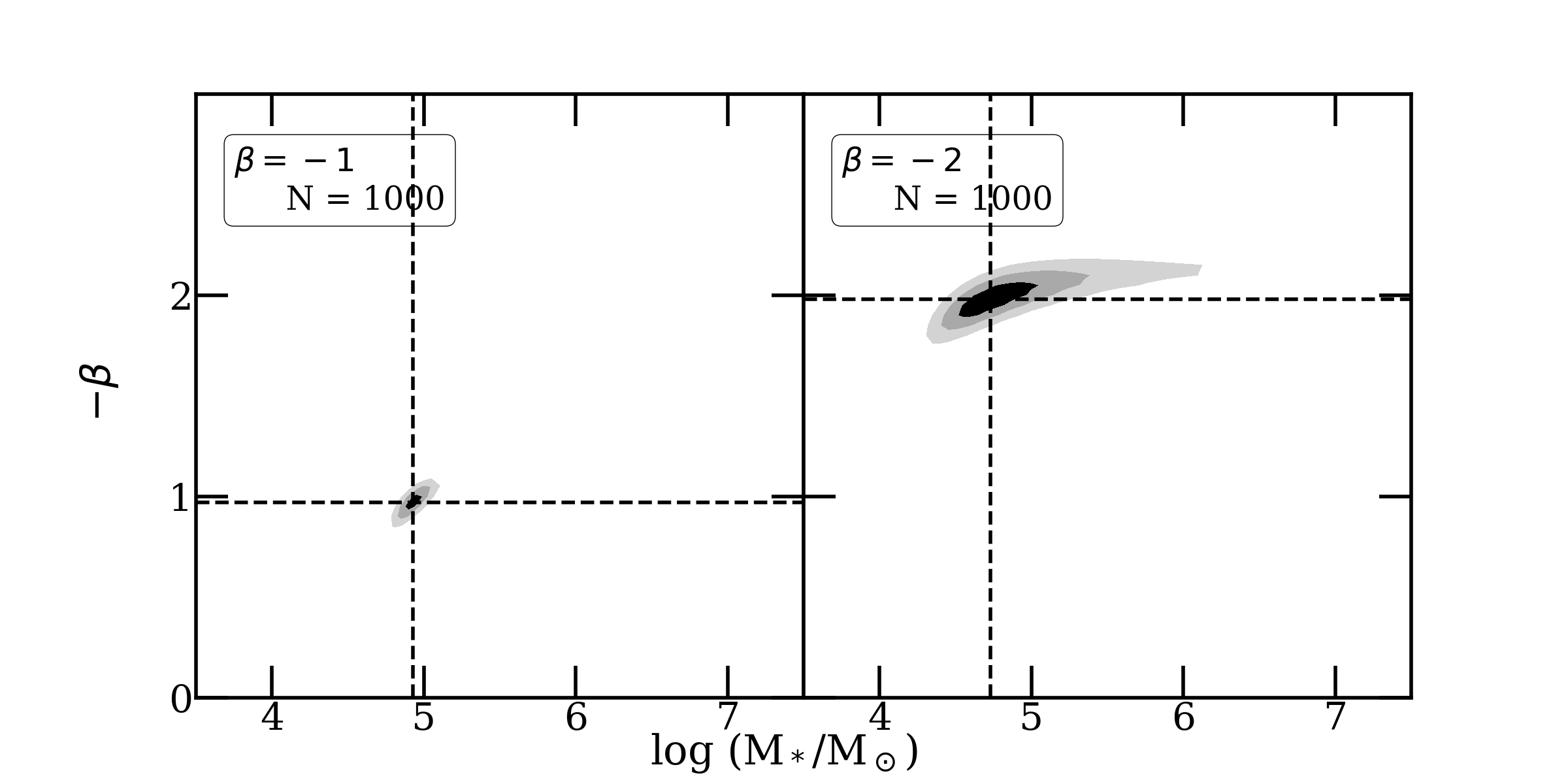}
	\caption{Our maximum-likelihood fitting method applied to mock cluster catalogs. We generate 1000 mock cluster members from an underlying Schechter distribution with $\beta = -1$ (left) and $\beta = -2$ (right), with $M_* = 10^5~M_\odot$. Note the significantly larger uncertainties in the $\beta=-2$ case, because of the difficulty of measuring a cutoff in steeper distributions.}
	\label{fig-appendix_2}
\end{figure*}

Finally, we test our ability to detect an upper mass cutoff for distributions with the same $M_*=10^5~M_{\odot}$, but different power law indices: $\beta=-1$ (such as for galaxies) vs. $\beta=-2$ (such as for clusters and GMCs). The results in Figure~\ref{fig-appendix_2} are based on generating 1000 simulated objects from each distribution, and performing the maximum-likelihood fitting described on each case. The smaller contours for the $\beta=-1$ case indicate that, for a given sample size, it is easier to detect a physical upper cutoff in populations with shallower power-law indices, such as found for for galaxies. The steeper $\beta\approx-2$ power-law index found for GMC and cluster populations makes the detection of an upper mass cutoff more challenging.
\end{document}